\documentclass{nature_sub}

\usepackage{graphicx}
\usepackage{amsmath}
\usepackage{amssymb}
\usepackage{color}
\usepackage{hyperref}
\usepackage{verbatim}

\newcommand{\nat}{{Nature}}

\newcommand{\blue}{\textcolor{blue}}
\newcommand{\red}{\textcolor{red}}
\def\red#1{\textcolor{red}{\textbf{}}}
\def\blue#1{#1}

\usepackage{multibib}

\newcites{main,meth}%
         {,%
          }
\bibliographystylemain{naturemag}
\bibliographystylemeth{naturemag}

\title{Stream-Disk Shocks as the Origins of Peak Light in Tidal Disruption Events}

\author{Elad Steinberg$^1, $Nicholas C. Stone$^{1}$}

\begin{document}

\maketitle

\begin{affiliations}
 \item Racah Institute of Physics, The Hebrew University, Jerusalem, 91904, Israel
\end{affiliations}

\begin{abstract}
Tidal disruption events occur when stars are ripped apart\cite{Hills75, Rees88} by massive black holes, and result in highly luminous, multi-wavelength flares \cite{KomossaBade99, Gezari+06, vanVelzen+11}. \blue{Optical/UV observations\cite{vanVelzen+11, Gezari+12, Arcavi+14} of tidal disruption events (TDEs) contradict simple models of TDE emission\cite{Rees88, Ulmer99}, but the debate between alternative models (e.g. shock power\cite{Piran+15, Shiokawa+15} or reprocessed accretion power\cite{LoebUlmer97, Guillochon+14, CoughlinBegelman14, MetzgerStone16, Roth+16, RothKasen18}) remains unsettled,} 
\red{Unfortunately, the geometry and power source\cite{LoebUlmer97, Piran+15} for flare optical/UV photospheres remain unclear,} as the dynamic range of the problem has so far prevented {\it ab initio} hydrodynamical simulations\cite{Lodato+20}.  Consequently, past simulations have resorted to unrealistic parameter choices\cite{Guillochon+14, Shiokawa+15, Hayasaki+13, Hayasaki+16, Bonnerot+16, Sadowski+16}, artificial mass injection schemes\cite{BonnerotLu20, Bonnerot+21}, or very short run-times\cite{Andalman+22}.  Here we present \blue{a}\red{the first ever }3D radiation-hydrodynamic simulation of \red{such} a \blue{TDE} flare from disruption to peak emission, with typical astrophysical parameters.  At early times, shocks near pericenter power the light curve and a novel source of X-ray emission, but circularization and outflows are inefficient. Near peak light, stream-disk \red{interactions}\blue{shocks} efficiently circularize returning debris, power stronger outflows, and reproduce observed peak optical/UV luminosities \cite{Hung+17, vanVelzen+21}.  Peak emission in this simulation is shock-powered, but upper limits on accretion power become competitive near peak light as circularization runs away.  This simulation shows how deterministic predictions of TDE light curves and spectra can be calculated \red{before the next generation of time-domain surveys, such as VRO and {\it ULTRASAT}, expands our sample from dozens\cite{vanVelzen+21} to thousands\cite{BricmanGomboc20, Jonker+20, BenAmi+22} of observed flares.}\blue{using moving-mesh hydrodynamics algorithms.}
\end{abstract}



\red{Optical/UV observations\cite{vanVelzen+11, Gezari+12, Arcavi+14} of tidal disruption events (TDEs) contradict simple models of TDE emission\cite{Rees88, Ulmer99}, but the debate between alternative models (e.g. shock power\cite{Piran+15, Shiokawa+15} or reprocessed accretion power\cite{LoebUlmer97, Guillochon+14, CoughlinBegelman14, MetzgerStone16, Roth+16, RothKasen18}) remains unsettled due to the lack of first-principles simulations.}  We have performed \red{the first }\blue{a }self-consistent simulation\red{s} of mass return and electromagnetic emission from \blue{a} TDE\red{s} with realistic and typical parameters: \blue{black hole mass} $M_\bullet = 1\times 10^6 M_\odot$, stellar mass $M_\star = M_\odot$, and pericenter $R_{\rm p}=R_{\rm t}$ (the tidal radius).  To simulate a ``typical'' TDE, we employ \red{a modified}\blue{an updated} version of \textsc{RICH}, a publicly available moving-mesh hydrodynamics code\cite{Yalinewich+15}.  \textsc{RICH} was recently modified to include a dynamical treatment of grey radiation transport (flux-limited diffusion) along with more realistic equations of state \blue{(including ionization and recombination)}, both of which are key elements of the TDE problem.  The gravitational field is a pseudo-Newtonian Paczyński–Wiita potential that approximates general relativity.  We additionally smooth the gravitational potential in its inner regions to reduce runtime.  We discuss these approximations further, and perform post-hoc self-consistency checks, in the Methods.  We run the simulation past the peak of the mass fallback rate (59 days), up until a final time of 65 days, roughly when the optical/UV light curve peaks.  We approximate the emitted thermal radiation in post-processing.

\begin{figure}
\begin{tabular}{cc}
\includegraphics*[width=8.5cm]{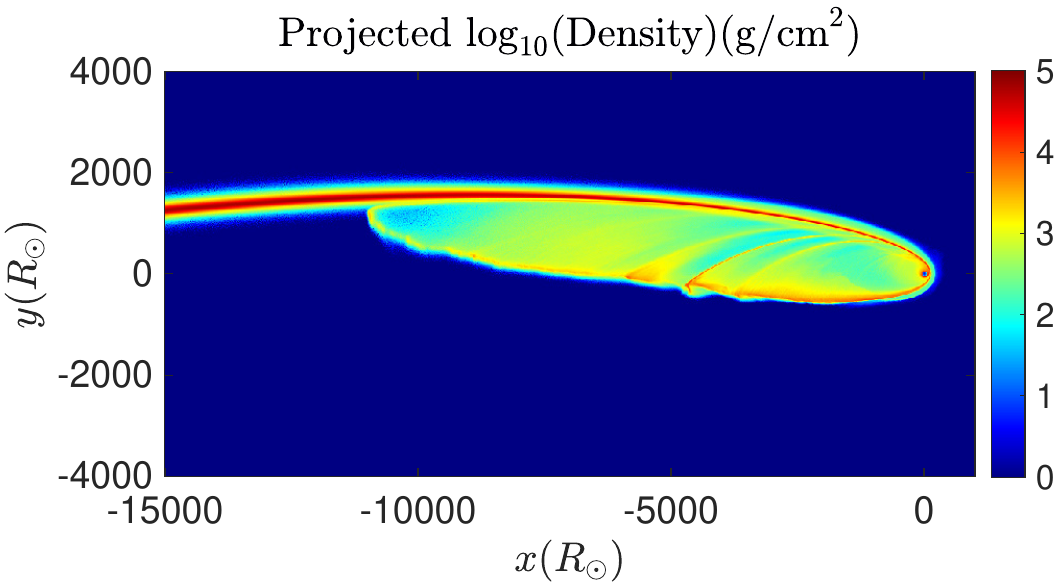} & 
\includegraphics*[width=8.5cm]{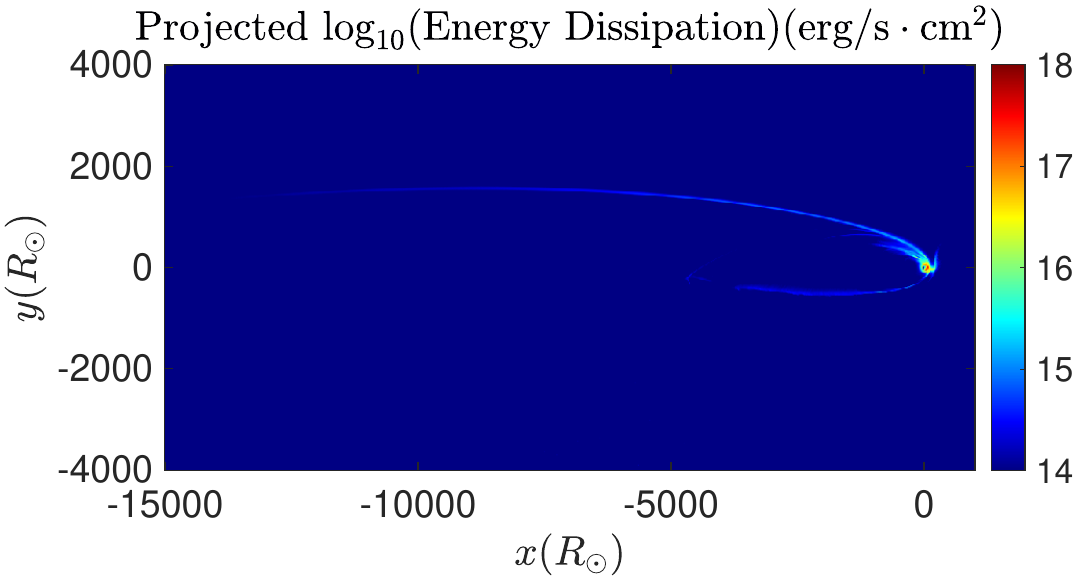} \\
\includegraphics*[width=8.5cm]{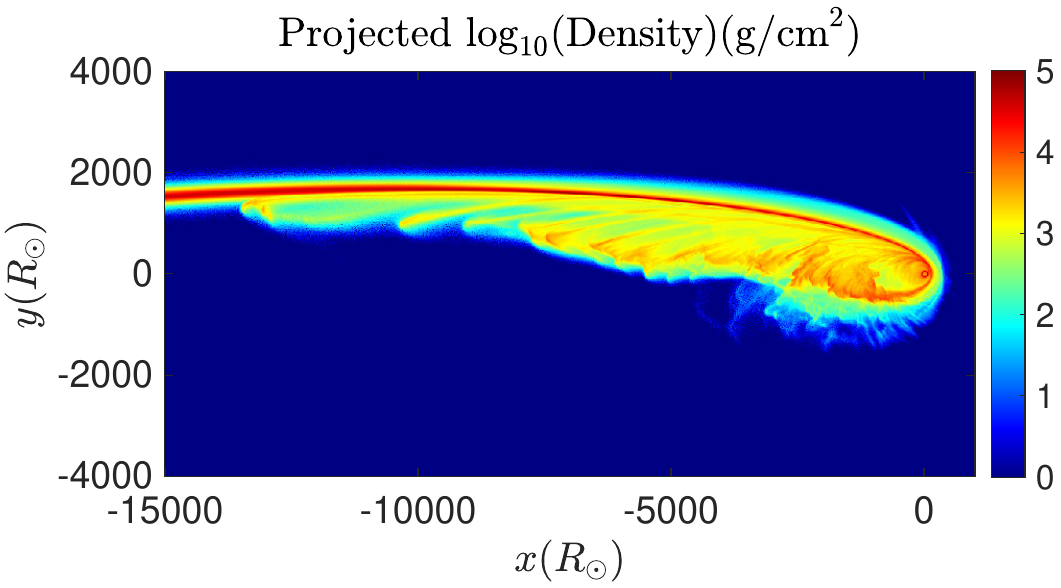} & 
\includegraphics*[width=8.5cm]{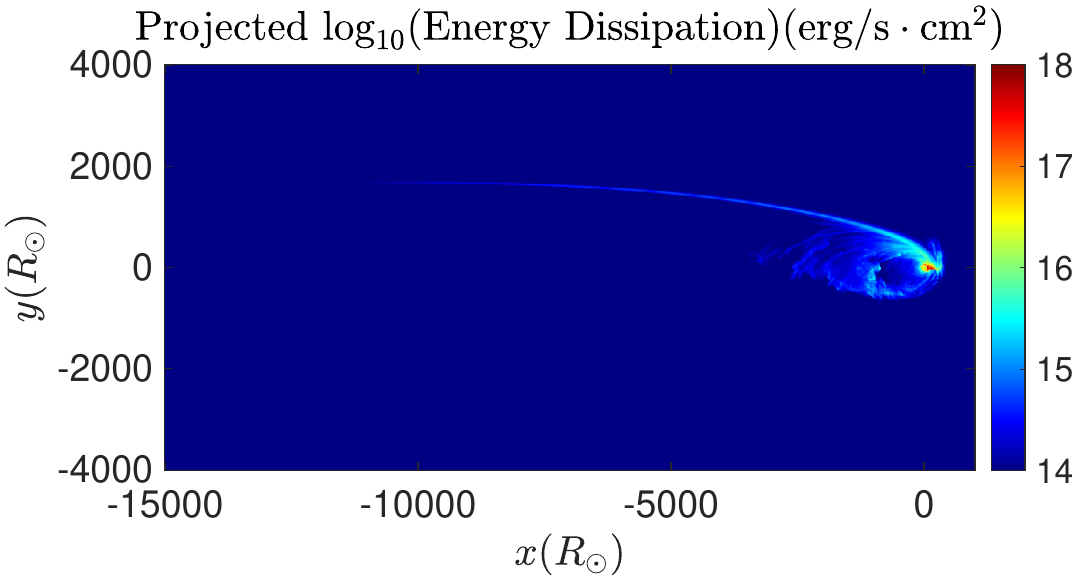} \\
\includegraphics*[width=8.5cm]{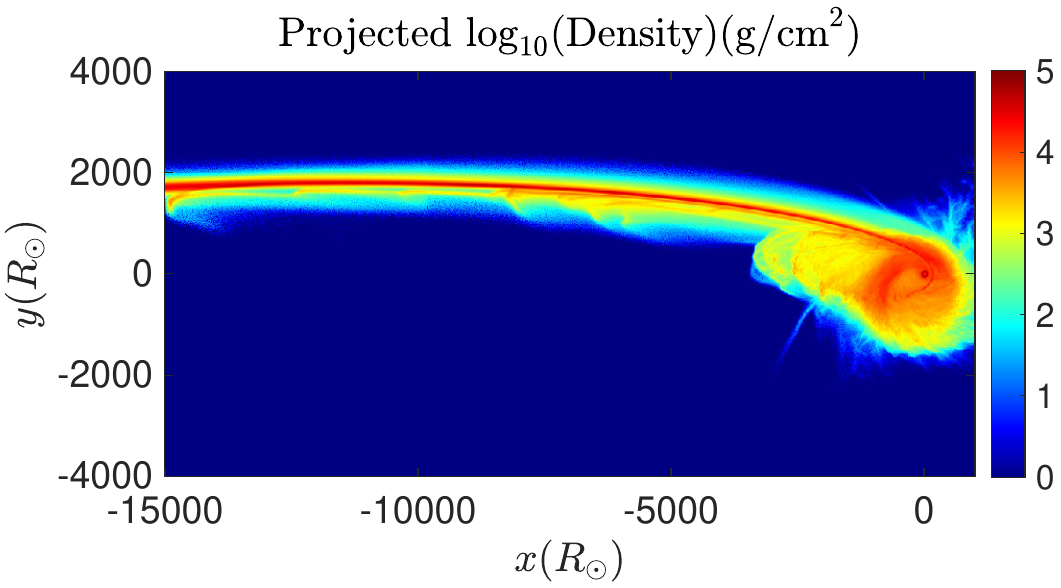} & 
\includegraphics*[width=8.5cm]{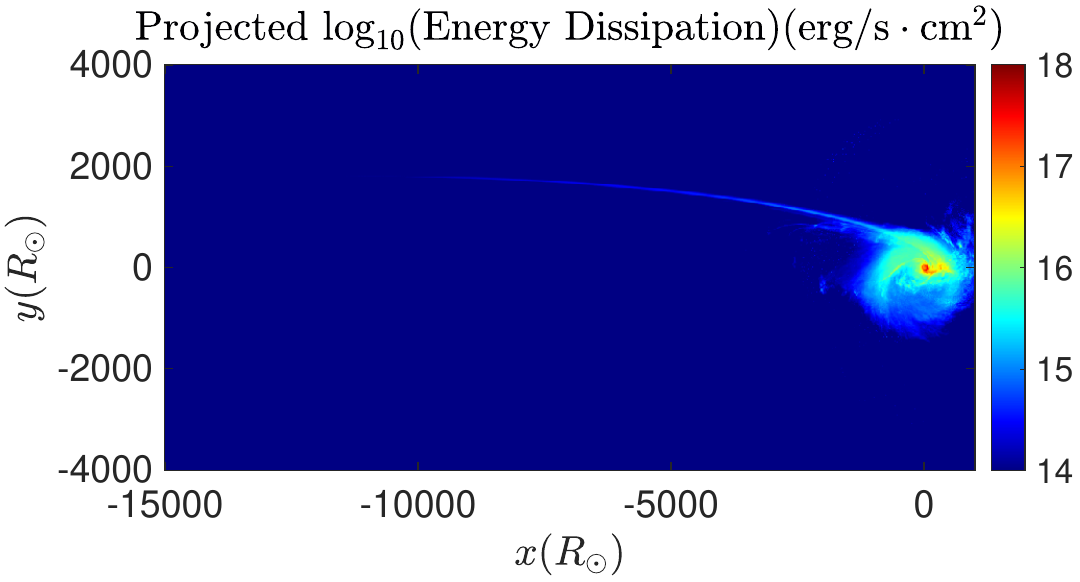} \\
\end{tabular}
\caption{{\bf Density and dissipation.}  Projections of the gas density $\rho$ (left) and volumetric dissipation rate $\dot{u}$ (right) on the orbital plane at $t=47$ (top), $t=55$ (middle), and $t=62$ (bottom) days.  The color schemes are logarithmic and are labeled in sidebars.  The $\dot{u}$ figures show that at all times, energy dissipation is dominated by interactions relatively near stream pericenter (rather than at the self-intersection radius visible in the density plots).  Comparing $t=47$ to $t=62$ day figures, we see qualitative morphological differences in both $\rho$ and $\dot{u}$.  At early times, dynamically cold streams produce the majority of their $\dot{u}$ in a small region localized near the compression shock at pericenter.  At late times, a runaway circularization process has begun, and $\dot{u}$ is spread over a larger region in which returning streams shock against partially circularized debris.
}
\label{fig:projection}
\end{figure}

Fig. \ref{fig:projection} shows snapshots of our results (see the Methods for more details): 2D projections of both gas density $\rho$ and the volumetric energy dissipation rate $\dot{u}$, at times $t=47$, $t=55$, and $t=62$ days post-disruption.  
The dominant sources of energy dissipation at early times are shocks associated with compression near pericenter, which are strong enough to begin the circularization process but too weak to efficiently circularize debris. At $t=47$ days, a small minority of the debris receives large energy perturbations (visible in the $t=47$ day density plot as a region ``filling in'' the nearly Keplerian ellipse of the debris stream).  This small fraction of the debris plays an outsized role in driving circularization.  By $t=62$ days, the morphology of both the debris and the dissipation zone has changed dramatically.

Inspection of the late-time density field in Fig. \ref{fig:projection} shows that the post-pericenter debris has increased its binding energy by roughly one order of magnitude.  The partially circularized debris is now impacted by streams returning for the first time, which efficiently dissipate their kinetic energy in shocks spanning a wide range of radii.  The transition from early-time dissipation (near the compression point) to late-time dissipation (from stream-disk interactions) shows that circularization can be a runaway process: the circularization of even a small amount of bound debris is enough to enhance the circularization efficiency for streams that have yet to return. 

\begin{figure}
\centering
\includegraphics[width=16.0cm]{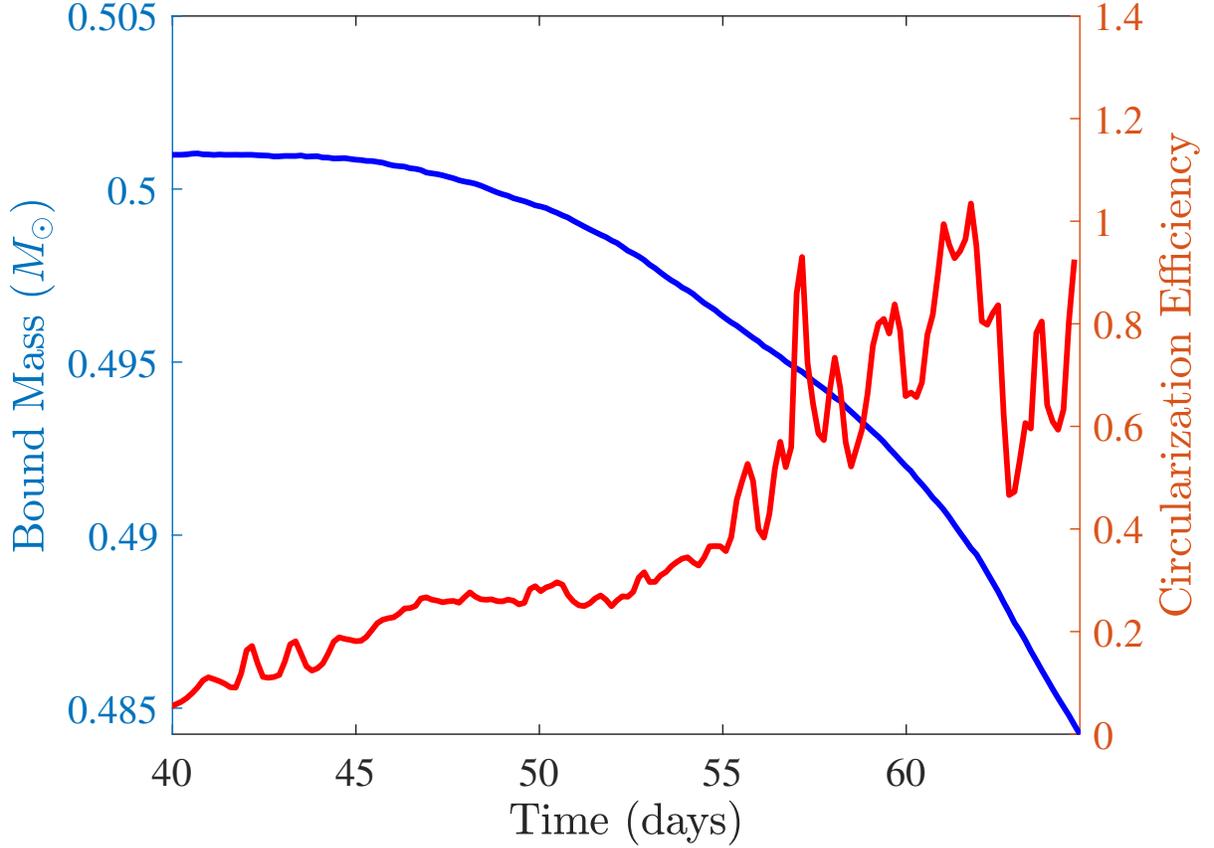}
\caption{{\bf Dynamical evolution of the bound debris.}  The blue curve shows the bound mass as a function of time.  While only a small minority ($\approx 3\%$) of the dynamically bound debris is unbound in outflows by the end of the simulation, this is $\approx 15\%$ of all gas that has passed once through pericenter. Red shows the dimensionless circularization efficiency $\chi(t)$.  If $\chi$ (Eq. \ref{eq:chi}) equaled $1$, all matter returning for the first time would instantly circularize.  Values of $\chi \ll 1$, as seen at early times, imply inefficient circularization. The steady rise in $\chi$ culminates in circularization becoming efficient near the end of the simulation ($\approx 57-65$ d).  }
\label{fig:circ}
\end{figure}

Fig. \ref{fig:circ} quantifies the shift in dissipation sites by calculating the time evolution of gas orbital energy $E$ (for bound mass only).  Using the dynamical mass fallback rate $\dot{M}_{\rm fb}(t)$ (see Methods), we translate the orbital energy dissipation rate $\dot{E}$ into a  dimensionless circularization efficiency,
\begin{equation}
    \chi = \frac{-\dot{E}}{GM_\bullet \dot{M}_{\rm fb}(t)/4 R_{\rm p}}.
    \label{eq:chi}
\end{equation}
The denominator is the dissipation rate required to force freshly returning debris to circularize immediately at fixed angular momentum.  Fig. \ref{fig:circ} shows that $\chi$ steadily increases during the simulation, beginning at very low values ($\chi \sim 0.01$) and eventually reaching high efficiencies ($\chi \approx 0.5-1$), as was suggested by eye in Fig. \ref{fig:projection}.

Fig. \ref{fig:circ} also quantifies the rate of mass loss in outflows, which have been seen in many past circularization simulations\cite{Sadowski+16, Bonnerot+21, Andalman+22}, although not in others\cite{Hayasaki+13, Shiokawa+15, Bonnerot+16} In our simulation, only $\approx 3\%$ of the bound mass is ejected.
While this is a small fraction of the bound mass, it is $\approx 15\%$ of all mass that has returned through pericenter, and the ejection efficiency steadily increases through the run.  By the end of the simulation, typical outflow speeds at infinity are $\approx 7500~{\rm km~s}^{-1}$, about an order of magnitude lower than predicted winds from super-Eddington accretion disks\cite{Sadowski+15}.

We calculate the bolometric lightcurve with two different methods. First, we sum the radiative flux through each photospheric cell given by our grey radiation transfer scheme.  Second, we approximate the emitted spectrum by computing the thermalization radius $R_{\rm th}$, and assuming that each cell above and below $R_{\rm th}$ emits a blackbody spectrum with its local temperature $T$, and a fractional weight given by the effective optical depth in this cell.  We then convert this approximate spectrum into a synthetic observation by calculating its luminosity in ZTF and {\it Swift UVOT} bands, and fitting a single-temperature blackbody to the ``multi-band photometry.''  Both procedures are done for 192 isotropically-sampled viewing angles.

\begin{figure}
\centering
\includegraphics[width=16.0cm]{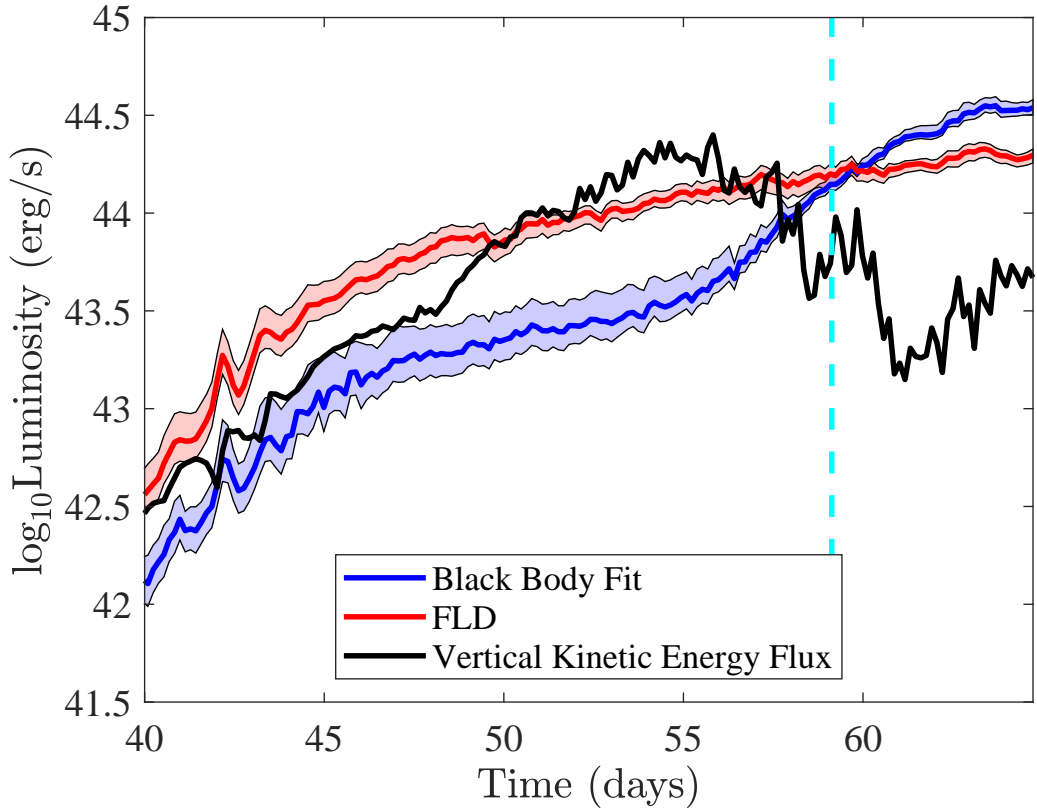}
\caption{{\bf TDE light curves.}  The self-consistent bolometric luminosity as a function of time (red), the fitted single-temperature black body luminosity $L_{\rm BB}$ (blue), and the flux of vertical kinetic energy passing through pericenter (black), all plotted against time since disruption.  The red and blue lines represent isotropic averages (over 192 isotropically sampled angles), while the shaded regions are $1\sigma$ contours.  The vertical kinetic energy flux initially tracks both luminosity measurements, highlighting its role in early-time emission, but decouples and declines later on (as stream-disk interactions become the dominant dissipation mechanism).  The dashed vertical line shows the peak of the dynamical mass fallback rate, slightly preceding peak light.}
\label{fig:L_fits}
\end{figure}

In Fig. \ref{fig:L_fits}, both luminosity estimators indicate that the simulation has roughly reached the peak of the light curve.  The bolometric/flux-limited diffusion approach estimates a substantially lower (higher) peak (pre-peak) luminosity than the single-temperature black body fit, likely due to the inherent inaccuracy of the latter method.  The flux of vertical kinetic energy (a proxy for energy dissipation in shocks produced by the compression and re-expansion of debris streams moving through the pericentric nozzle) roughly tracks the flux-limited diffusion luminosity until a time $\approx 55$ days, when the vertical kinetic energy flux enters a sharp decline.  This is further evidence for a qualitative transition from early-time dissipation in compression shocks to late-time dissipation from runaway stream-disk interactions.  The peak ``fitted'' luminosity exhibits only a weak dependence on viewing angle.

\begin{figure}
\begin{tabular}{cc}
(a)\includegraphics[width=8.0cm]{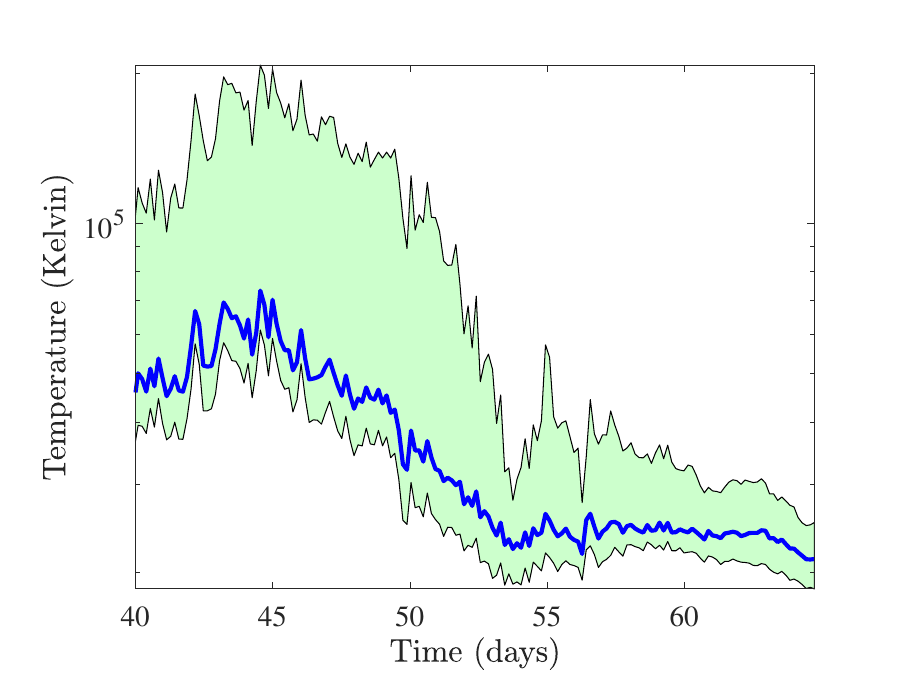} & 
(b)\includegraphics[width=8.0cm]{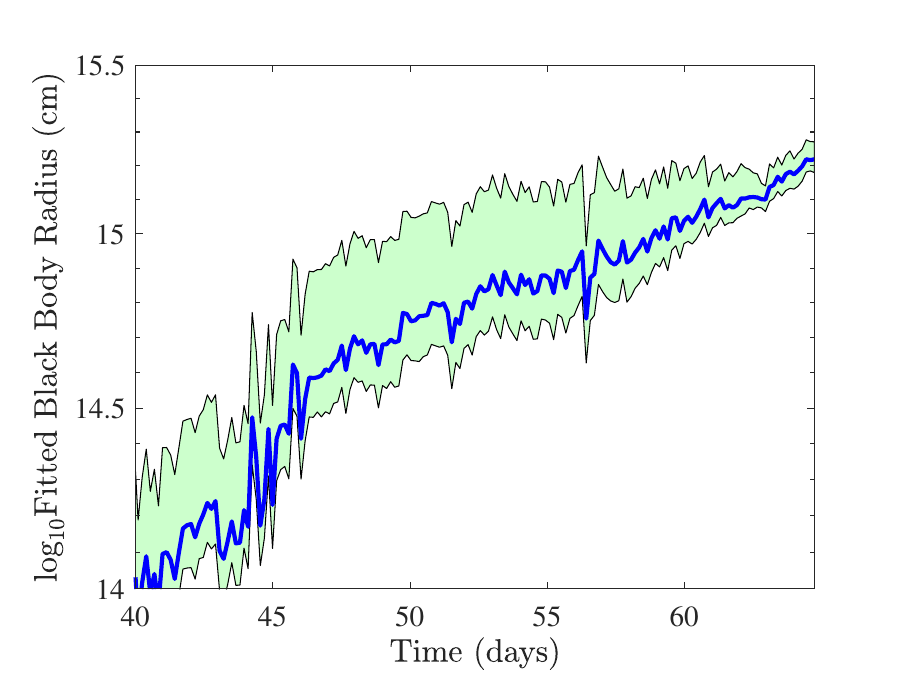} \\
(c)\includegraphics[width=8.0cm]{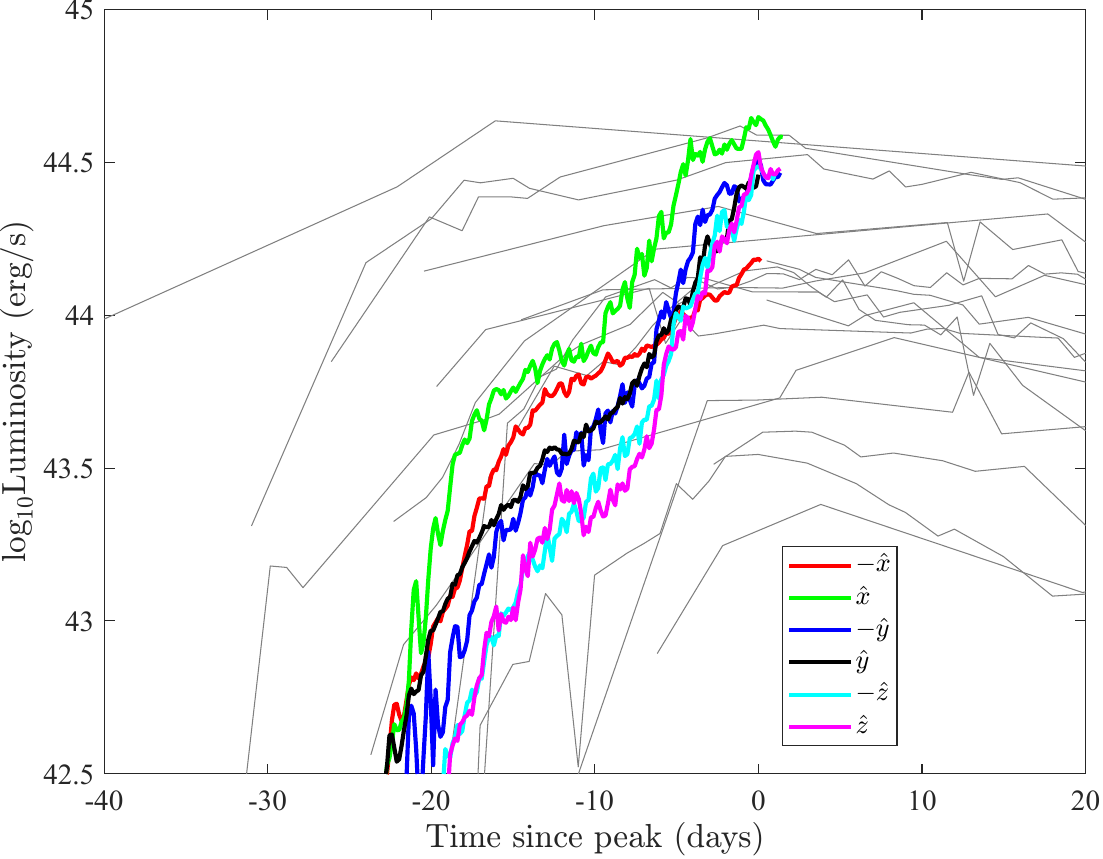} & 
(d) \includegraphics[width=8.0cm]{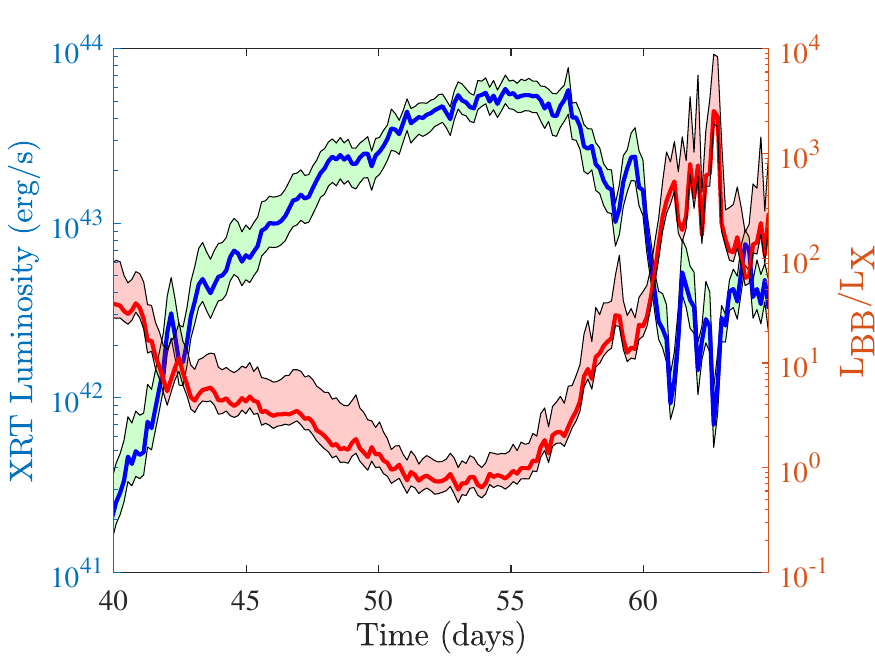} \\
\end{tabular}
\caption{{\bf Observables as a function of time.}  {\it Top left}: the fitted blackbody temperature considered over a range of isotropically sampled observer viewing angles.  The blue solid line is the average temperature evolution, while the green shaded region shows the $1\sigma$ contour.  {\it Top right}: the fitted blackbody radius as a function of time, with line and shading the same as before.  {\it Bottom left}: fitted blackbody luminosities from the simulation's six principal axes (the star's final orbit is in the $xy$-plane, and pericenter is at $y=z=0$, $x=R_{\rm p}$) shown for comparison against 17 ZTF TDE light curves\cite{vanVelzen+21}.  {\it Bottom right}: {\it Swift XRT}-band X-ray luminosities $L_{\rm X}$ (blue) and $1\sigma$ angular variation (green), the synthetic optical/X-ray ratio $L_{\rm BB}/L_{\rm X}$ (dark red), and angular $1\sigma$ variation (light red).}
\label{fig:obs}
\end{figure}

Fig. \ref{fig:obs} shows the fitted blackbody parameters as a function of time: blackbody temperature $T_{\rm BB}$, blackbody luminosity $L_{\rm BB}$, and the derived blackbody radius $R_{\rm BB}$ (assuming simultaneous {\it Swift} and ZTF observations).  We also compare our blackbody light curves to a sample of 17 real TDE light curves observed by ZTF\cite{vanVelzen+21}.  Empirically, a large fraction of TDEs feature powerful optical/UV emission from extended photospheres \cite{Hung+17, vanVelzen+21}, with peak $L_{\rm BB} \sim 10^{43 - 44.5}~{\rm erg~s}^{-1}$, $R_{\rm BB} \sim 10^{14-15}~{\rm cm}$ at peak (with significant time evolution before/after peak), and $T_{\rm BB} \approx 2-5 \times 10^4$ K at peak (with little time evolution before/after peak).  Fig. \ref{fig:obs} shows generally excellent agreement between our simulation's results and this qualitative range of observed blackbody parameters near the peak of the TDE light curve. In the Methods, we compare $R_{\rm BB}$ to physical radii of interest and find that it is a poor proxy for \blue{both} dissipation regions and the color surface.

Insofar as our synthetic, fitted blackbody light curves are anomalous in any way, it is that they feature a relatively steep rise to peak, a relatively high $L_{\rm BB}$, and a relatively large $R_{\rm BB}$ at peak.  However, none of these features are outside the range of what is seen in observed TDEs\cite{Hung+17, vanVelzen+21}.  We speculate that the large values of $L_{\rm BB}$ and $R_{\rm BB}$ produced in our simulation may be due to our choice to simulate the full disruption of a $M_\star = 1M_\odot$ star.  Partial disruptions and low-mass stars make up the majority of all TDEs\cite{StoneMetzger16}, and the lower mass fallback rates of these events would naturally be expected to reduce $L_{\rm BB}$ and $R_{\rm BB}$.

In the Methods, we drop our single-temperature blackbody fitting procedure and compute the full multi-color blackbody spectrum, $\nu L_\nu$, as seen at a range of viewing angles and a range of times.  Notably, there is significant quasi-thermal emission at soft X-ray wavelengths, {\it powered primarily from compression shocks}, before any significant circularization has occurred.  We quantify this novel X-ray emission in panel (d) of Fig. \ref{fig:obs}, where we estimate the total luminosity $L_{\rm X}$ produced in the {\it Swift} XRT bandpass.  We find a peak angle-averaged X-ray luminosity of $\approx 10^{43.5}~{\rm erg~s}^{-1}$, which is achieved around days 50-57.  After peak, the X-ray light curve enters into a steep decline due to the decreasing efficiency of the pericenter shocks (e.g., compare the synthetic {\it Swift} XRT light curve to the vertical kinetic energy flux in Fig. \ref{fig:L_fits}) and the concomitantly increased efficiency of circularization, which enshrouds the (now weakened) pericentric shocks in a larger photosphere that does not allow X-rays to escape.  TDE X-ray luminosities of $\approx 10^{43.5}~{\rm erg~s}^{-1}$ are in principle observable, but $L_{\rm X}$ peaks 1-2 weeks prior to maximum optical light, posing an observational challenge.  For example, while the ZTF sample features extensive {\it Swift} X-ray coverage, it does not include X-ray observations of any flares two weeks prior to peak optical light\cite{vanVelzen+21}.

There are several limitations to our approach that we will address in future work.  First, our treatment of gravity is both pseudo-Newtonian and uses an artificially softened potential at small radii for reasons of computational efficiency.  While our post-hoc checks demonstrate that these approximations do not strongly affect the results (see Methods), a more realistic treatment of gravity would be needed for more relativistic TDEs, or for longer-term evolution of the TDE we have simulated.  The softened potential means that we can only place an upper limit on accretion luminosity, and while this limit is sub-dominant to shock power at peak light, it is of the same order of magnitude, so we cannot firmly exclude that accretion may play a competitive role.  Second, our work does not include magnetic fields, although past simulations have found that MHD does not strongly affect circularization dynamics\cite{Sadowski+16}.  Third, our treatment of radiation transport is approximate, being both grey and diffusive.  Our post-processed spectra are approximate and should be studied in the future with full Monte Carlo radiative transport.  Finally, and most importantly, this paper presents the simulation of only a single TDE, with specific values of $M_\bullet$, $M_\star$, and $R_{\rm p}$.

\red{Despite these limitations, this is still the first ever {\it ab initio} TDE simulation to produce an optical/UV light curve that reaches peak light, for an astrophysically realistic TDE.  Furthermore, it is one of only a handful of long-term TDE simulations to include a dynamical treatment of radiation and the first to include a realistic equation of state involving hydrogen ionization/recombination, which plays a surprisingly important role in stream hydrodynamics and dissipation (see Methods).  Our results qualitatively resemble some previous generations of TDE simulations, but with key differences.  For example, w}\blue{W}hile we agree with some past simulations\cite{Shiokawa+15} that shock power can initially dominate over accretion, our results differ in that circularization has become efficient by peak light.  Likewise, our finding of runaway circularization through stream-disk shocks recalls some recent works\cite{Bonnerot+21, Andalman+22}, although in contrast, the ``self-intersection'' shock is largely irrelevant, and circularization is kick-started by the pericenter shock.  \red{Unlike all these past simulations, we have chosen typical TDE parameters and avoided artificial mass injection schemes.  }Crucially, there are important qualitative differences between our results and all past approximate simulation schemes, emphasizing the importance of realistic parameter choices in future TDE simulations.  The observational rate of TDE detection is poised to increase by 1-2 orders of magnitude over the next few years\cite{BricmanGomboc20, Jonker+20, BenAmi+22}, but turning detections into measurements of MBH demographics\cite{StoneMetzger16}, nuclear stellar dynamics\cite{StoneMetzger16}, and fundamental physics\cite{Lu+17, Wen+21} will require broader parameter studies in radiation-hydrodynamic TDE simulations.



\begin{addendum}
 \item[Acknowledgments] ES and NCS thank Assaf Horesh, Peter Jonker, Brian Metzger, Tsvi Piran, and Elena Maria Rossi for helpful discussions.  ES and NCS also thank Clement Bonnerot, Richard Saxton, and especially the two anonymous referees for their constructive feedback. NCS received support from the Israel Science Foundation (Individual Research Grant 2565/19) and the BSF portion of a NSF-BSF joint research grant (NSF grant No. AST-2009255 / BSF grant No. 201977.
  \item[Author Contributions] ES ran the simulations of the TDE and produced the figures.  He also modified the publicly-available RICH code to include radiation transport and a realistic equation of state.  NCS wrote the paper.  Both authors contributed equally to the physical interpretation of the numerical results.
\end{addendum}

\newpage

\clearpage



\begin{methods}

TDEs were once discovered as exotic one-off events\cite{Bade+96, KomossaBade99, Gezari+06}, but are now found at ever increasing rates by wide-field time-domain transient surveys\cite{Arcavi+14, Hung+17, vanVelzen+20, vanVelzen+21}, usually due to supermassive black holes (SMBHs) in galactic nuclei.  Tidal disruption of main sequence stars presents a superficially simple problem.  In the Newtonian limit, the leading-order outcome is dictated by just three variables: black hole mass $M_\bullet$, stellar mass $M_\star$, and orbital pericenter $R_{\rm p}$ (the orbit is usually parabolic to very high precision\cite{StoneMetzger16}).  Here we have assumed that knowledge of $M_\star$ is sufficient to determine the stellar radius and internal structure, variables which also play a role in the disruption process\cite{Lodato+09}.  Since most TDEs come from main sequence stars, this assumption is approximately correct, but will be modified modestly by age along the main sequence\cite{GallegosGarcia+18, LawSmith+20}.  If general relativity is important (i.e. if $R_{\rm p} \lesssim 10 R_{\rm g}$, where the gravitational radius $R_{\rm g} = GM_\bullet / c^2$; note however that the exact number of gravitational radii separating a ``relativistic'' from a ``non-relativistic'' pericenter in the spectrum of possible TDEs is highly uncertain at present), then the MBH spin $\chi_\bullet$ and spin-orbit misalignment angle $\iota$ will matter also\cite{GuillochonRamirezRuiz15, Hayasaki+16}.  The small number of controlling parameters is the reason why TDEs are believed to hold great promise for measuring SMBH demographics ($M_\bullet$, $\chi_\bullet$).  However, this potential is fundamentally limited by the lack of first-principles models for the non-standard accretion flows produced in the aftermath of tidal disruption.

Although past simulations have generally worked in astrophysically unrealistic limits of the TDE problem, there is much that can be learned from them.  Simulations of TDEs around intermediate-mass black holes (IMBHs) universally fail to exhibit rapid circularization, both in Newtonian\cite{Guillochon+14} and general relativistic\cite{Shiokawa+15} gravity.  This may be a consequence of the non-relativistic orbital pericenter ($R_{\rm p}/R_{\rm g} \sim 100$), and results in the formation of a long-lived, globally eccentric accretion flow.  As little material makes it to the black hole event horizon, the primary early-time power source in these accretion flows is from shocks\cite{Shiokawa+15}.  Conversely, the ``eccentric TDE'' limit of simulations finds that SMBH TDEs can circularize efficiently due to relativistic precession, so long as $R_{\rm p}/R_{\rm g}$ is small enough\cite{Hayasaki+13}.  In this limit, circularization may be delayed by a large pericenter\cite{Bonnerot+16} or alternatively by misaligned SMBH spin\cite{Hayasaki+16}.  How to ``interpolate'' between these two non-astrophysical limits is not obvious, if it is possible at all.  More recent works have simulated TDEs with astrophysically realistic parameters, but at the expense of using artificial mass injection schemes\cite{BonnerotLu20, Bonnerot+21} or only running for a handful of days after the start of mass fallback\cite{Andalman+22}.  The former approach is based on clever but ultimately uncertain idealizations that in any case are likely to break down after some amount of time (as the motivating local simulations\cite{LuBonnerot20} of stream-stream intersections assume that streams pass through pericenter without any interactions with material that has accumulated at small radii), while the latter approach, while rigorous, has not yet succeeded at making predictions for the brightest parts of a TDE light curve.  This bottleneck in first-principles TDE simulation is the fundamental problem motivating this paper.

In this Methods, we provide greater detail on our numerical methods (\S \ref{app:numerics}), the numerical convergence of our results (\S \ref{app:converge}), the evolution of tidally disrupted debris streams (\S \ref{app:stream}), the process of circularization and energy dissipation (\S \ref{app:circ}), and our approximate calculation of the emergent spectra of TDEs near peak light (\S \ref{app:observe}).  Finally, we compare our results to a broad survey of the past simulation literature (\S \ref{sec:comparison}).

\section{Numerical methods}
\label{app:numerics}

We initialize our simulation by modeling the star as an $n=3/2$ polytrope with mass $M_\star = M_\odot$ and radius $R_\star = R_\odot$.  The star's volume is initially filled with approximately $2\times 10^7$ cells, and its center of mass is placed at a distance $r = 3R_{\rm t}$ away from a SMBH with mass $M_\bullet=10^6\;M_\odot$, where the tidal radius $R_{\rm t} = R_\star (M_\bullet / M_\star)^{1/3}$.  The star is placed on a zero-energy orbit with pericenter $R_{\rm p} = R_{\rm t}$.  As mentioned in the main text, we use a pseudo-Newtonian potential that approximates relativistic apsidal precession\cite{PaczynskiWiita80}, but we smooth it as 
\begin{equation}
    {\bf F}=-\frac{GM_\bullet}{h\left(h-2R_{\rm g}\right)^2}{\bf r},
\end{equation}
when $r<h$, where we have chosen $h=60R_\odot$. This choice of smoothing was made for computational convenience, but our post-hoc checks in \S \ref{app:circ} show that it is unlikely to affect our main results.  Throughout the simulation, gas self-gravity is calculated using a quadrupole moment tree code with an opening angle $\theta=0.9$.  

We use a realistic equation of state (EOS) that assumes equilibrium abundances of ionized, neutral and molecular hydrogen including the rotational
and vibrational degrees of freedom, and neutral, singly- and
doubly-ionized helium\cite{Tomida+15, Pejcha+16}. The composition is X = 0.7, Y = 0.3,
and the ratio of ortho- to para-H$_2$ is assumed to be in
thermodynamical equilibrium with the gas temperature. The latent heats for all processes are consistently included.


\textsc{RICH} solves the radiation transport problem with a flux-limited gray diffusion scheme using the Larsen flux limiter, where the diffusion coefficient $D$ is
\begin{equation}
    D=c\left(\left(3\kappa\rho\right)^n+\left(\frac{|\nabla E_r|}{E_r}\right)^n\right)^{-1/n}.
\end{equation}
Here $n=2$, $\kappa$ is the Rosseland mean opacity, $c$ is the speed of light, $E_{\rm r}$ is the radiation energy density and $\rho$ is the gas density.  
We assume Thomson scattering for the scattering opacity $\kappa_s = 0.34 \textrm{ cm}^2/\textrm{g}$. For the absorption opacity $\kappa_{\rm a}$, we use values calculated from \textsc{cloudy} version 17.02\cite{Ferland+17}, assuming solar abundances and local thermodynamic equilibrium (LTE).  We continue to assume LTE to transform these into values for emission, i.e. $\rho\kappa_a aT^4$, as discussed in more detail in \S \ref{app:observe}.  

We begin with standard equations for radiation transport\cite{Krumholz+07}, but slightly alter them in two ways. First, we modify the advection of the radiation energy term, to take into account the movement of the mesh. Second, we multiply the coupling term between the radiation and the gas, i.e. $\rho\kappa_a c \left(E_r - aT^4\right)$, by the Fleck factor\cite{Fleck+71}, $f$,
defined as
\begin{equation}
    f=\frac{1}{1+c\beta_f\rho\kappa_a \Delta t}.
\end{equation}
Here $\Delta t$ is the time step, $\beta_f=\partial (aT^4)/\partial e_{\rm th}$, $c$ is the speed of light, and $e_{\rm th}$ is the thermal energy per unit volume. This reduces the stiffness of the coupling term by making the thermal energy equation semi-implicit.  Solving the diffusive radiation transport equations self-consistently determines gray radiation variables such as $E_{\rm r}$.

Initially, the simulation is run in the center of mass frame of the star, in order to reduce the overall magnitude of local velocities and thus numerical advection errors. Once the tip of the returning gas reaches a distance of $r=10^3R_\odot$, this reference frame is no longer beneficial, and we switch to a stationary lab frame. This switch is a simple Galilean transformation of the velocity and introduces no numerical error.  In order to achieve high resolution near pericenter, we use an adaptive mass resolution scheme. Once we switch to the lab frame (around day 20) and gas starts falling back towards the SMBH, we maintain a minimum (maximum) mass resolution of $0.5 \times 10^{-7} M_\odot$ ($1.5 \times 10^{-7} M_\odot$) per cell for cells far from pericenter and decrease it quadratically with distance from the SMBH for regions with $r<3\cdot10^3R_\odot$, i.e, $m=10^{-7} M_\odot\left(\frac{r}{3\cdot10^3R_\odot}\right)^2$ until a cell reaches a minimum spatial extent of $1\;R_\odot$. At the end of the simulation, the total number of cells is $\sim 2\times 10^8$.

\section{Convergence Study}
\label{app:converge}
In this section we examine the convergence of our simulation with numerical resolution. We are particularly interested in examining the outcome of vertical compression and the resulting ``nozzle shock'' as streams return to pericenter for the first time\cite{Kochanek94}, as (i) past simulations have found a wide range of outcomes of this shock, (ii) the amount of stream expansion from the nozzle determines the importance of later stream self-intersections\cite{BonnerotStone21, BonnerotLu22}, and (iii) this aspect of the TDE problem is the most challenging to resolve\cite{BonnerotLu22}.  Initially, returning streams are focused to a caustic-like point of maximum compression near pericenter, akin to the original compression of the disrupted star\cite{CarterLuminet83, Stone+13} (in both cases, compression is driven by free fall of tidally disrupted fluid in the quasi-homologous vertical tidal field).  Portions of the stream initially closer to the orbital plane experience an adiabatic bounce, reversing their vertical compression, but are then impacted by infalling material that began its collapse from a larger height; this impact creates an (approximately) standing shock in the lab frame\cite{Shiokawa+15}.  Past global simulations of TDEs around IMBHs have found significant dissipation in the nozzle region, leading to a wide-angle spray of material exiting its first pericenter return\cite{Guillochon+14, Shiokawa+15}.  For global simulations of TDEs around SMBHs, the picture is less clear, and the amount of dissipation seems to depend on a combination of the underlying equation of state\cite{Hayasaki+13, Bonnerot+16} and spin-orbit misalignment\cite{Hayasaki+16, Liptai+19}; numerical resolution and stellar eccentricity are likely to affect the outcome as well.

Recent 2D simulations, performed in an idealized transverse slice of a debris stream (moving in the non-inertial frame of the stream's center of mass), have examined the problem of the nozzle shock with much higher vertical resolution than can be achieved in a 3D simulation\cite{BonnerotLu22}.  The returning gas streams undergo a drastic compression, accompanied by significant shock dissipation.  However, by the time the post-shock debris returns to the self-intersection radius, the cylindrical widths and densities are comparable to the width of the gas stream that is infalling for the first time\cite{BonnerotLu22}. This implies a high efficiency of dissipation at the self-intersection radius $R_{\rm SI}$, which is not seen in our simulations and thus merits careful examination. 

We investigate the importance of the nozzle shock for later  dissipation at $r\approx R_{\rm SI}$ by running our own array of comparable 2D simulations using the 2D version of \textsc{RICH}.  First, we re-simulate the 2D problem studied by Bonnerot \& Lu (hereafter BL22\cite{BonnerotLu22}), employing the full non-linear tidal force of a Paczyński–Wiita potential, an adiabatic ideal gas EOS with an adiabatic index of 5/3, and the initial conditions of BL22. As seen in ED Fig. \ref{fig:convergence} (panel a), if we initialize our simulations with the BL22 initial conditions, our result is similar to theirs, both in terms of peak compression and in the final height (and density) once the post-shock stream returns to its initial starting distance.  At this later point in time, the width of our 2D stream is slightly larger than the one  in BL22 (by a factor of $\lesssim 2$), due to the fact that we used the full non-linear tidal force, the non-homologous components of which modestly alter the collapse and rebound\cite{Stone+13}. Next, we check the effect of choosing the initial temperature to be extremely cold, by redoing the previous simulation, but with a more realistic initial stream temperature of 4000K (compare to 70K in BL22). The resulting height of the stream at the end of this simulation is about a factor of 3 greater than in the simulation with the original BL22 initial conditions (ED Fig. \ref{fig:convergence}).

We now check how the 2D slice evolution changes if the stream is started from initial conditions taken directly from our 3D simulation, which is both hotter (4000K, as already simulated) and more dilute (by an order of magnitude, partially due to recombination energy injected by the EOS) than the initial conditions in BL22. These initial conditions corresponds to a gas element with a specific orbital energy of $-94GM_\odot/R_\odot$, initialized at a distance of $r=5120R_\odot$ from the SMBH.  The stream is then evolved inward through pericenter and back out until it reaches the self intersection point, at a distance of $r \approx 8000R_\odot$. We use the same (realistic) EOS as in the 3D simulation. The yellow line in ED Fig. \ref{fig:convergence} (panel a) shows that while the stream is greatly compressed at pericenter, by the time it returns to $R_{\rm SI}$ there is roughly a factor of 3.5 
increase in its height relative to that of the infalling streams.
This mismatch in scale heights (and thus densities) between outgoing and ingoing streams naturally explains the very low dissipation rate that is observed in our 3D simulation.

In our 3D simulation, the vertical resolution at pericenter is much poorer than what can be achieved using idealised 2D runs, and we address this by redoing the last, most realistic 2D simulation with a reduced resolution that is comparable to that in our 3D simulation. The compression achieved at pericenter is far less than what is seen in the previous 2D simulations due to the lack of grid cells to resolve the vertical collapse. However, this ultimately does not have a large effect on the key dynamical results, as the final height of the outgoing stream at the self-intersection point is less than a factor of 2 greater than in the realistic, high resolution 2D run (compare the outer terminus of the yellow solid and purple dashed lines in ED Fig. \ref{fig:convergence}, panel a).  In both the high-resolution and the low-resolution simulations, the stream height (density) at the self-intersection point has increased (decreased) by at least a factor of $\sim 10^{1.5}$ ($\sim 10^3$) compared to our reproduction of the BL22 results.  This difference is primarily due to the combination of (i) more realistic -- namely, hotter and more diffuse -- initial conditions for the stream, and (ii) a realistic EOS (which in addition to altering the initial conditions also injects late-time heat into the outgoing stream).  In realistic 3D simulations, the limited sensitivity to resolution that we see in our 2D runs may be further decreased by added sources of dissipation, such as 3D shearing motions that are absent in 2D transverse slices.  In summary, the comparison of our high- and low-resolution 2D runs shows that while our global 3D simulation certainly underresolves peak stream compression, this does not change the qualitative nature of post-pericenter stream expansion, which is the ultimate explanation for the lack of a strong self-intersection shock.

Aside from these 2D convergence studies, we further examined the dynamics of stream compression and the nozzle shock with a short-term 3D re-simulation that has a factor of 5 higher global resolution.  This simulation is initiated when the most tightly bound debris passes apocenter and stops at day 28, after the most tightly bound debris has passed through pericenter.  The post-pericenter gas density field is very similar between our fiducial run and this high-resolution re-simulation, providing further supporting evidence that the key qualitative outcome of the nozzle shock (a relatively wide-angle stream exiting pericenter) is reasonably converged.

We have focused our attention so far on how vertical resolution impacts the nozzle shock, as this is by far the most challenging aspect of TDE debris circularization to fully resolve. However, the global convergence of our simulation is, of course, also important. We investigate this by re-running our entire 3D simulation with a reduced resolution, featuring a factor of 5 decrease in the total number of cells. ED Fig. \ref{fig:convergence} (panel b) shows the bolometric lightcurve (extracted directly from the flux-limited diffusion of the radiation field) for our nominal run and also for the low-resolution run. At the earliest times, when the least bound gas (which is most poorly resolved) infalls, we see that our low-resolution run substantially overpredicts the emission, due to increased numerical dissipation.  The two runs have converged to better than a factor of $2$ by $t=45$ days.  By the time that the novel soft X-ray emission has reached peak ($t\approx 55$ days), the two runs have converged to within a few tens of percent, and by the time of peak bolometric light ($t\approx 65$ days), convergence is excellent.  As it is these later times which produce detectable levels of radiation, we consider our global predictions to be acceptably converged.

\section{Stream evolution}
\label{app:stream}

Following disruption, we start by calculating the mass fallback rate at two different epochs, in both cases estimating it as $\dot{M}_{\rm fb} = ({\rm d}M/{\rm d}\epsilon)({\rm d}\epsilon/{\rm d}t)$.  The first epoch is at $t=0.4$ days, when the debris center of mass is at a distance $r \approx 10 r_{\rm t}$ from the SMBH, and the second is just before the first mass falls back to pericenter, at a time $t=23.17$ days.
Extended Data (ED) Fig. \ref{fig:lambda_radius} (panel a) shows that during these first $\approx$ 20 days, the imputed fallback rate changes only negligibly (the overall orbital energy changes by less than a percent), demonstrating (i) that any energy injection via recombination fails to immediately alter the leading-order orbital dynamics\cite{Coughlin+16}, (ii) a lack of fragmentation due to stream self-gravity (in contrast to some recent simulations\cite{Hayasaki+20}), and (iii) that the code conserves energy.  

The first of these points can be understood physically by comparing the specific orbital energy shift imparted from recombination, $\delta \epsilon_{\rm rc} \sim \sqrt{\epsilon_{\rm rc}} v_{\rm orb}(r)$, to the actual specific orbital energy of the debris, which for the most tightly bound debris is $|\epsilon| \sim \Delta\epsilon = GM_\bullet R_\star / R_{\rm t}^2$, the frozen-in energy spread\cite{Stone+13, Steinberg+19}.  Here $\delta \epsilon_{\rm rc} \sim 10~{\rm eV}/m_{\rm p}$ is the energy injected per unit mass by electron recombination.  As the ratio $\delta \epsilon_{\rm rc} / \Delta \epsilon \sim 10^{-2}$, we only should expect a very minor perturbation to the mass fallback rate from recombination.  Notably, we do not see any gravitational fragmentation of the debris streams in our simulation, as is generically expected for sufficiently stiff EOSs\cite{Coughlin+16} and has specifically been seen in recent simulations that incorporate ionization and dissociation effects\cite{Hayasaki+20}.  This discrepancy in results could be due to the different stellar models used ($n=3$ versus $n=3/2$ polytropes), or perhaps due to artificial surface tension in smoothed-particle hydrodynamics techniques.  

Although the effect of recombination on bulk stream energetics is minor, it plays a larger role in the initial dissipation mechanisms that operate in our simulation.  As discussed in the main text, the compression of returning streams leads to shock heating that powers the initial light curve (on timescales $\lesssim 55$ days since disruption).  However, the  specific kinetic energy of this vertical collapse will be determined by the scale height of the streams, $H$, at the time they themselves tidally disrupt\cite{Guillochon+14,Coughlin+16}.  By puffing up the streams, recombination heating creates a greater reservoir of gravitational potential energy to be dissipated during the vertical collapse process, increasing the efficiency of this power source compared to simple estimates.  We have quantified this by re-running the early stages of our simulation with a simpler, adiabatic EOS.  The comparison between our fiducial run (with the realistic EOS) and the adiabatic EOS is shown in ED Fig. \ref{fig:lambda_radius}.  Panel (b) of this figure compares the debris stream linear density $\lambda$ between these two runs, and finds little difference (unsurprisingly, as the linear density is set by total energetics, and as already discussed, $\delta \epsilon_{\rm rc} \ll \Delta \epsilon$).  However, in panel (c), we calculate and present the cylindrical radius that encloses various percentages of the stream mass in the plane orthogonal to its local bulk velocity vector.  As this cylindrical radius is a good proxy for the vertical scale height $H$, we refer to it as ``height'' here. We find a factor $\approx 3$ increase in stream scale height near apocenter (the rough point of stream tidal disruption) when recombination is included in the EOS.  This will enhance the efficiency of shocks near pericenter, increasing the total energy budget available for this early-time luminosity source.  Since vertical collapse velocity scales roughly linearly with initial height of a stream, this suggests that recombination may add roughly one order of magnitude to the kinetic energy of vertical collapse, compared to an adiabatic EOS.

In contrast with some past work \cite{KasenRamirezRuiz10}, we finds that the debris streams are too optically thick to radiate the energy injected via H recombination.  Considering the outgoing stream leaving pericenter for the first time, we find that the ionization fraction drops to $50\%$ once the stream reaches a distance $r_{\rm rec} \approx 1500 R_\odot$, at a density of $\rho \approx 1\times 10^{-4}~{\rm g~cm}^{-3}$.  Assuming (conservatively) a relatively narrow stream width of $H = 1 R_\odot$, we estimate a minimum stream optical depth of  $\tau \sim \rho H \kappa_{\rm es} \sim 3 \times 10^6$.  This is a minimum optical depth at recombination because we have taken the opacity to be electron scattering-only, $\kappa = \kappa_{\rm es} = 0.40~{\rm cm}^2~{\rm g}^{-1}$; in reality, bound-free opacities may be significantly greater.  We can now use this minimum optical depth at recombination to estimate a minimum photon diffusion time at recombination, $t_{\rm diff} \sim H \tau / c \sim 74 ~{\rm d}$.  Notably, this timescale is roughly two orders of magnitude larger than the dynamical time at recombination, $t_{\rm dyn} = \sqrt{r_{\rm rec}^3 / (GM)} \sim 1~{\rm d}$, and almost three orders of magnitude longer than the characteristic stream expansion time $t_{\rm exp} = H / c_{\rm s} \sim 0.3~{\rm d}$ (here we have taken a characteristic sound speed at recombination $c_{\rm s}\sim 30~{\rm km~s}^{-1}$).  We also note that even this long photon diffusion time is only a lower limit on the true stream cooling time, as the streams are generally dominated by gas rather than radiation pressure.

In summary, any adiabatic work done by the energy injected by recombination appears very likely to expand the stream before suffering radiative losses.  When examining the ionization fraction $x$, we see smooth gradients of $x$ along the stream's axial and transverse directions, rather than a sharp recombination front characteristic of optically thin material.  Some past analytic work\cite{KasenRamirezRuiz10} on TDE stream recombination assumed homologous stream expansion, which rapidly reduces stream densities and optical depths and thus produces a regime where radiative losses are important; in contrast, our simulation finds that stream self-gravity makes the hydrodynamic expansion far from homologous.

\section{Circularization}
\label{app:circ}



As we discuss in \S \ref{sec:comparison}, this simulation is in many ways the most astrophysically realistic realization of TDE circularization and evolution to date.  We have taken astrophysically typical parameters ($M_\bullet = 10^6 M_\odot$, $M_\star = M_\odot$, $\beta=1$, $e=1$) and have not employed artificial mass injection schemes.  However, our approach still suffers from a number of limitations, including a lack of magnetic fields, an approximate treatment of radiation transfer, and of course the restriction to a single set of parameters.  The largest limitation of our simulation is likely our treatment of gravity, which is approximate in two ways: first, the Paczyński–Wiita (PW) approximation of relativistic precession, and second, an artificial softening of the potential at small radii, as mentioned in \S \ref{app:numerics}.  The first of these approximations does not seriously reduce the realism of our results, as the standard TDE parameters we have selected feature only a weakly relativistic pericenter.  The importance of relativistic apsidal precession is determined by the self-intersection radius\cite{Dai+15} of a debris element with pericenter $R_{\rm p}=R_{\rm t}/\beta$ and eccentricity $e$:
\begin{equation}
    R_{\rm SI} = \frac{(1+e)R_{\rm t}}{\beta(1-e\cos(\delta \omega /2))}.\label{eq:RSI}
\end{equation}
Here the apsidal shift angle $\delta \omega$ is the amount of precession that occurs per orbit (and we have assumed here that relativistic precession occurs impulsively at pericenter passage).  At leading post-Newtonian (PN) order, the apsidal shift is
\begin{equation}
    \delta\omega_{\rm 1PN}= \frac{6\pi GM_\bullet}{R_{\rm p}(1+e)}, 
\end{equation}
while in the PW potential, the apsidal shift asymptotes to a value of 
\begin{equation}
    \delta\omega_{\rm PW}= \frac{2}{3} \delta \omega_{\rm 1PN}
\end{equation}
for the highly eccentric orbits relevant to us\cite{Wegg12}.  For the parameters in our simulation, $R_{\rm p} \approx 47 R_{\rm g}$, and the most tightly bound debris streams have initial $e\approx 0.99$.  At leading post-Newtonian order, Eq. \ref{eq:RSI} predicts $R_{\rm SI} \approx 6.2\times 10^3 R_{\rm g}$, while the PW potential produces $R_{\rm SI} \approx 7.7 \times 10^3 R_{\rm g}$, a difference of $\approx 20 \%$.  We have already seen by eye in Fig. \ref{fig:projection} that dissipation at $R_{\rm SI}$ remains highly subdominant throughout the course of our simulation, a qualitative point that we will quantify later in this section.  The fundamental reason for this subdominance is not the $\approx 20 \%$ error that the PW potential produces in $R_{\rm SI}$, but rather the large mismatch between densities of ingoing and outgoing streams at $R_{\rm SI}$, as can be seen in both the 2D and 3D simulations we show in ED Fig. \ref{fig:convergence}.  Consequently, we do not expect that the choice of PW potential has a significant impact on our results, though it probably would for a more relativistic TDE ($R_{\rm p} \sim R_{\rm g}$).

The second of our approximations may be of larger importance: to reduce computational expense, we have chosen a softening length of $h=60R_\odot$ (this is akin to the extended inner boundary employed by some past simulations\cite{Shiokawa+15}).  This approximation is potentially problematic: if a large amount of mass accumulates at these small radii, it would be able to generate substantial radiative or mechanical luminosity that could couple to the material at larger scales.  

We perform a post-hoc check on whether this assumption has affected our results in ED Fig. \ref{fig:Mr}, where we show the amount of mass that is accumulated within different radial cuts.  As we see from this figure, only a small amount of matter is ever present in radii $r \le h$; by the end of the simulation at $t\approx 65$ days, roughly $M_{\rm sm}\approx 2\times 10^{-3} M_\odot$ is enclosed within the smoothing radius.  If we assume that $100\%$ of this matter accreted with a radiative efficiency of $\eta=0.05$, then its time-averaged accretion luminosity over 65 days would be $L_{\rm acc} \approx  10^{43.5}~{\rm erg~s}^{-1}$, almost an order of magnitude less than our final peak luminosity of $\approx 10^{44.3}~{\rm erg~s}^{-1}$.  We have further estimated the time-dependent accretion luminosity as $L_{\rm acc} = \eta \dot{M}_{\rm acc} c^2$, by taking the net rate of mass entry into the $r\le h$ region and defining this as $\dot{M}_{\rm acc}$.  This gives a somewhat higher value of $L_{\rm acc}$ at peak, $\approx 1\times 10^{44}~{\rm erg~s}^{-1}$, which is smaller than the shock-powered luminosity in the simulation, but only at the factor of 2 level.  However, this approximate accretion luminosity is likely an upper limit for three reasons.  First, at these near-Eddington values of $\dot{M}_{\rm acc}$, significant amounts of thermal energy can be advected into the SMBH horizon, reducing $\eta$ below the fiducial Schwarzschild value we have taken.  Second, the short viscous times at small radii mean the actual gas density at $r \le h$ will be far lower than in our simulation, hampering the ability of small radii to capture high-$e$ stream material directly through shock dissipation.  Finally, the lower optical depths in the vertical direction will enable much of the accretion luminosity to escape without being reprocessed into optical/UV bands, likely making the accretion flow a subdominant source at these wavelengths. We ultimately conclude that at peak light, accretion power is subdominant {\it or comparable to} shock power.  As our upper limit on accretion power reaches the same order of magnitude as the shock luminosity, we cannot draw a stronger conclusion.

In summary, our neglect of orbital dynamics at small radii is not affecting our leading-order outcomes, but is nonetheless a weakness that may alter some quantitative conclusions.  This neglect would not be appropriate for a more relativistic TDE (either a high-$\beta$ disruption around a $10^6 M_\odot$ SMBH, or a low-$\beta$ disruption around a larger SMBH), and it will likewise become inadequate at later times for the TDE we have simulated, when more stellar debris eventually dissipates enough energy to circularize.  However, it appears reasonable for timescales corresponding to the peak of the optical/UV light curve in typical TDEs, as we have simulated here.

In the main text, we have presented various lines of evidence suggesting a gradual transition between two different dissipation zones.  At early times, dissipation appears (by examination of Fig. \ref{fig:projection}) to be dominated by internal shocks near and slightly downstream from the point of maximum compression at pericenter, akin to the nozzle shock seen to play a large role in IMBH TDEs\cite{Shiokawa+15}.  This early-time dissipation site is relatively weak, leading to circularization efficiencies $\chi \lesssim 0.2$.  However, at late times (after $t \approx 50-55$ days), the locus of dissipation shifts to a larger, more spatially extended region, apparently driven by shocks between returning debris streams and an eccentric disk of partially circularized material.  

Here we develop this approximate picture in greater detail by examining the geometric features of dissipation in our simulation.  First, we will examine the morphology of the debris density and dissipation zones by eye, and then we will use more quantitative metrics.  In ED Fig. \ref{fig:densityGeometry}, we show projected gas densities $\rho$ in the $xz$ and $yz$ planes.  We choose three snapshots in time which are the same as the times of the $xy$ density projections in Fig. \ref{fig:projection}.  With these ``side views,'' we can see a morphological transition.  At early times (e.g. the $47$ day snapshot), the $xz$ and $yz$ projections are dominated by a dynamically cold, geometrically thin stream of returning debris.  Little material has exited pericenter (i.e. moved to negative $y$ values), and the material that has retains a small aspect ratio.  By late times (e.g. the $62$ day snapshot), a large, quasi-ellipsoidal bound accretion flow has formed around the SMBH, with a pronounced $+y/-y$ asymmetry driven by interactions between the returning streams and the newborn flow.  We note that the $yz$ plane projections feature a low density region at the most negative $y$ values; this is the distant, receding material from the unbound arm of the tidal debris.

 In ED Fig. \ref{fig:dissipationGeometry}, we show projections of the volumetric dissipation rate $\dot{u}$ onto the $xz$ and $yz$ planes.  As before, we choose three snapshots in time which are the same as the times of the $xy$ projections in Fig. \ref{fig:projection}.  Once again, we see a transition in the basic morphology of the dissipation loci as time progresses.  Initially, dissipation is confined to a narrow (and very geometrically thin) range of $xy$ coordinates corresponding to the vicinity of the compression point.  By $t=62$ days, the locus of dissipation has expanded greatly in the $x$ direction, and moderately in the $y$ and $z$ directions, indicating a transition in shock locations.

We now quantify this transition in ED Fig. \ref{fig:dissipationR} (panel a), which presents the time evolution of the dissipation radius, $r_{\rm diss}=\int r \dot{E}dV/\int\dot{E}dV$.  This radius, which represents an average spatial scale on which dissipation occurs, is initially quasi-static at $r_{\rm diss} \approx 150 R_\odot$.  Around day $45$ it begins a slow expansion to $r_{\rm diss} \approx 350 R_\odot$, where it remains until day $\approx 55$, after which it begins a rapid expansion, reaching values $r_{\rm diss} \approx 1000 R_\odot$ by the peak of the optical light curve.  This rapid growth in the dissipation radius near peak light reflects the formation of an extended, eccentric accretion flow, which experiences both internal shocks (akin to the ``secondary shocks'' seen in some recent work\cite{BonnerotLu20}) and also shocks from collisions with debris streams returning for the first time.

We see further evidence for this shift in dissipation zones in ED Fig. \ref{fig:dissipationR} (panel b), where we plot characteristics of the azimuthal angles $\phi$ where dissipation occurs.  The angle $\phi$ is defined in the orbital, or $xy$, plane.  Positive $y$ (or, equivalently, positive $\phi$) describes points that are {\it upstream of pericenter}, and negative $y$ (or negative $\phi$) describes points {\it downstream of pericenter}.  We now define a mean dissipation vector, ${\bf R}_{\rm diss} =\int \hat{\bf r} \dot{E}dV/\int\dot{E}dV$, and use this to compute the mean azimuth of dissipation, $\cos\phi_{\rm diss} = {\bf R}_{\rm diss} \cdot \hat{\bf x} / |R_{\rm diss}|$ (here $\hat{\bf x}$ and $\hat{\bf r}$ are unit vectors along the $x$-axis and the radial direction of a fluid element, respectively).  Note that $R_{\rm diss} = |{\bf R}_{\rm diss}| \le 1$ by construction, and in general the dimensionless $R_{\rm diss} \neq r_{\rm diss}$, which has units of physical distance.  The magnitude of the dissipation vector measures the azimuthal localization of the dissipation zone: $R_{\rm diss}=1$ implies that all dissipation is occuring at a single value of $\phi$, while $R_{\rm diss}=0$ implies axisymmetric dissipation rates.  The angle $\phi_{\rm diss}$ is generally positive at early times, oscillating roughly between $0$ and $0.4$ radians (regions upstream of the pericentric nozzle).  At early times, the shock dissipation zone is also moderately localized, with $R_{\rm diss}\approx 0.5$.  Starting around day $57$, $\phi_{\rm diss}$ begins to oscillate more violently, first becoming negative and then rapidly becoming positive.  At the same time, the shocks delocalize in azimuth, with $R_{\rm diss}$ falling to values $\approx 0.2$.

As a final metric to gauge the circularization of returning debris, we directly plot eccentricity evolution on a ``space-time'' diagram in Fig. \ref{fig:ecc_history}.  In this diagram, we bin up all grid cells around a certain radius and compute a mass-weighted mean eccentricity $\langle e \rangle$.  At all radii, we see a nearly monotonic decrease in eccentricity.  At $t=40$ days, all $r \gtrsim R_{\rm t}$ feature mean eccentricities $\langle e \rangle \gtrsim 0.9$, while at $t=65$ days, this is only true for radii $r\gtrsim 10^4 R_\odot = 100 R_{\rm t}$.  By the end of our simulation, substantial circularization has taken place within $r=1000 R_\odot = 10 R_{\rm t} \approx 500 R_{\rm g}$, with average eccentricities $\langle e \rangle \lesssim 0.5$.  While runtime limitations prevent us from self-consistently estimating the future evolution of the bound debris, we note that, at $t=65$ days, the pace of circularization is fastest at large, poorly circularized radii and has slowed at smaller, significantly circularized radii.  Likewise, while our smoothing of the inner potential prevents us from confidently predicting features of any inner accretion flow, a naive extrapolation of Fig. \ref{fig:ecc_history} suggests that low eccentricities may predominate inside $R_{\rm t}$.

Another important question in TDE dynamics is the rate of mass ejection in outflows from the system.  There are various theoretical reasons to consider strong outflows plausible: for many TDEs (including the one simulated in this paper), initial values of $\dot{M}_{\rm fb}$ are highly super-Eddington, and super-Eddington accretion disks are generally seen in simulations to unbind a large fraction of their mass in outflows\cite{Sadowski+15, Dai+18, Jiang+19}.  Even without any accretion, it is energetically feasible for the circularization process itself to eject a large fraction of the bound mass, due to its very low initial binding energy\cite{MetzgerStone16}.

As was already shown in Fig. \ref{fig:circ}, our simulation sees $\approx 3\%$ of the dynamically bound mass, or $\approx 15\%$ of the mass to have returned through pericenter, ejected in an outflow.  Typical outflow speeds at the end of the simulation are $\approx 7500~{\rm km~s}^{-1}\approx 0.025 c$.  This relatively low outflow velocity is about an order of magnitude smaller than typical disk winds from compact, super-Eddington accretion disks\cite{Sadowski+15, Dai+18, Jiang+19}, raising the prospect of shock dissipation at later times if matter begins to efficiently circularize at the smallest scales, launching a super-Eddington outflow that can overtake the slow-moving circularization outflow.

\section{Observables}
\label{app:observe}

We identify the location of the photosphere, $R_{\rm ph}$, at each snapshot by integrating radially inwards and equating
\begin{equation}
\int^\infty_{R_{\rm ph}}\rho(\kappa_{\rm a}+\kappa_{\rm s}){\rm d}r=\frac{2}{3}
\end{equation}
along 192 evenly spaced angles (here $\kappa_{\rm a}$ and $\kappa_{\rm s}$ are the Planck mean absorption and scattering opacities, respectively, although $\kappa_a\ll\kappa_s$ above the photosphere so only the scattering contributes to the photosphere location).  We find for each angle an inner thermalization radius $R_{\rm th}$ given by
\begin{equation}
\kappa_\textrm{eff}=\sqrt{3\kappa_a(\kappa_a+\kappa_s)},\quad\quad\tau_{\rm eff}=\int^\infty_{R_{\rm th}}\rho\kappa_\textrm{eff}{\rm d}r=1.
\end{equation}
We then approximate the emergent spectrum by assuming that each cell above and below $R_{\rm th}$ emits a blackbody spectrum.  The relative contribution for the \emph{i}th cell, at radius $r_i$, is given by
\begin{equation}
\rho_i\kappa_{a,i}aT^4_i\exp{\left(-\int_{r_i}^\infty\rho\kappa_\textrm{eff}{\rm d}r\right)}dV_i/\int^\infty_{R_{\rm th}}\rho\kappa_aaT^4 {\rm d}V, 
\end{equation}
and the overall normalization is given by the radiative flux through the photosphere.  We integrate down to an effective  optical depth $\tau_{\rm eff}=5$ along each of our 192 solid angles in order to compute the emissivity of each cell along each radial ray.  When calculating the spectral energy distribution (SED) seen along any individual line of sight, we downweight contributions from cells off the line of sight by a factor of $\cos\theta$.  This method is only a 1D approximation of a full Monte Carlo radiative transfer calculation, and in particular neglects lateral scatterings, but still gives a useful model for the overall features of the emitted spectrum. As mentioned in the main text, we mock up commonly used observational techniques by calculating the spectral luminosity in the ZTF bands $r,\;g$ and $i$, and in the \emph{Swift} bands $UVOT,\;UVW1,\;UVM2$ and $UVW2$. We then fit the final result in all bands to a single-temperature blackbody and derive the blackbody luminosity $L_{\rm BB}$, radius $R_{\rm BB}$ and temperature $T_{\rm BB}$.  

A key assumption in this calculation is that emissivity can be computed from tabulated \textsc{cloudy} absorption opacities $\kappa_{\rm a}$.  This assumption relies on LTE in the gas and radiation fields, and the tabulated $\kappa_{\rm a}$ values themselves are computed under the assumption of LTE.  As LTE will break down in sufficiently dilute and optically thin regions, we check the self-consistency of this assumption post-hoc by estimating the fraction of all luminosity coming from cells with radiation-gas equilibrium (quantified by checking whether the radiation and gas temperatures are equal to within $10\%$).  We find that $\approx 96\%$ of all emergent luminosity in our calculation comes from cells where the radiation and gas fields are in LTE, indicating that non-LTE effects likely play a subdominant role in setting the overall spectrum.

ED Fig. \ref{fig:rTherm} (panel a) shows the average (both arithmetic and geometric) values of $R_{\rm th}$ and $R_{\rm ph}$ as a function of time.  The thermodynamic properties of the gas are such that the photosphere radius is always outside the thermalization radius (i.e. scattering dominates absorption), which has important implications for emission line profiles\cite{RothKasen18}.  The behavior of $R_{\rm th}$ over time is non-trivial, and we explore it further here.  The large initial difference between these two averages shows that initially, the thermalization surface is highly asymmetric, but as time goes on and these two estimates converge, becomes more symmetric.  It is instructive here to compare characteristic radii in Fig. \ref{fig:obs}b, ED Fig. \ref{fig:dissipationR}, and ED Fig. \ref{fig:rTherm}.  At the end of the simulation, $R_{\rm BB}\sim 2\times 10^{15}$ cm, roughly one order of magnitude larger than $R_{\rm th}\sim 1-2\times 10^{14}$ cm and $R_{\rm diss}\sim 7 \times 10^{13}$ cm.  While $R_{\rm BB}$ is at least the same order of magnitude as $R_{\rm ph}$, it is still a factor $\sim 4-5$ larger, leading us to conclude that fitted single-temperature blackbody radii can be poor proxies for the size of regions where most dissipation occurs, the color radius of the TDE, and even its photospheric radius.  The fundamental reason for this mismatch is not aspherical geometry (note the lack of viewing angle dependence in Fig. \ref{fig:obs}), but rather the danger inherent in fitting multi-color thermal emission (especially emission peaking beyond the near-UV\cite{Arcavi22}) with a single-temperature blackbody.

ED Fig. \ref{fig:rTherm} (panel b) shows the dimensionless average radial velocity of the fluid cells that make up the thermalization surface -- specifically, the physical velocity $|V_{\rm r}|$ normalized by the local escape velocity $V_{\rm esc}$.  At all times, the thermalization surface is, on average, comprised of gravitationally bound material, although it appears to become more tenuously bound as time progresses.

We present the overall spectral energy distributions (SEDs) in ED Fig. \ref{fig:nuLnu}, at three different snapshots in time.  We see a general shift in the emergent spectrum towards lower energy photons as time goes on, broadly reflecting two of the primary trends in debris evolution we have emphasized (the delocalization of dissipation from a narrow region of compression shocks to a much broader region of stream-disk interactions, and also the expansion of the thermalization radius).  Soft X-ray emission is visible at all times and from all viewing angles, though in all cases the soft X-ray band is on the Wien tail of the overall spectrum, creating strong viewing angle dependence (particularly in comparison to optical/UV light).  These X-rays are generally emitted by low-density gas somewhat above $R_{\rm th}$.  As a sanity check, we recalculated these spectra under the assumption that $\tau_{\rm eff}<1$ cells emitted only via optically thin free-free radiation and found qualitatively similar soft X-ray spectra, suggesting that this prediction is qualitatively robust despite our approximate SED calculation.  However, this early-time X-ray signal should be better quantified in the future by full Monte Carlo ray-tracing calculations.  A small bump in the $t=45.73$ and $t=56.12$ spectra at hard X-ray wavelengths comes from a very small amount of very high-temperature gas far beyond $R_{\rm th}$ where our predictions are less robust, but the luminosities here are unobservably low.

As was already seen in Figs. \ref{fig:L_fits} and \ref{fig:obs}, we find that broadband photometric properties of the optical/UV photosphere have a surprisingly weak dependence on viewing angle, especially near the peak of the light curve.  We quantify this dependence more carefully in ED Fig. \ref{fig:T_hist}, where we present histograms showing the distribution of synthetic blackbody fit parameters $L_{\rm BB}$, $T_{\rm BB}$, and $R_{\rm BB}$.  These histograms are generated by isotropically sampling 192 different viewing angles.  We see that the fitted blackbody luminosity $L_{\rm BB}$ has a particularly narrow range of variation near peak light, which is promising for future efforts to estimate more fundamental parameters (e.g. $M_\bullet$) from broadband TDE light curves.

In Fig. \ref{fig:X_hist}, we explore analogous viewing-angle effects on the X-ray emission produced in our simulations.  We see that at $t=65$ days, the typical X-ray luminosity is $\sim 10^{42-43}~{\rm erg~s}^{-1}$, with an outlying tail of dimmer X-ray luminosities concentrated on the negative $x$-axis.  The optical-to-X-ray ratio $L_{\rm BB}/L_{\rm X}\sim 100$, much larger than its $\sim 1$ value 10-15 days earlier.  We suspect that this is the reason this soft X-ray precursor has not yet been identified: {\it Swift} XRT followup is generally obtained for TDEs {\it at the time of, or after}, peak light \cite{vanVelzen+20}.  Future observations that aim to identify the early X-ray transient predicted in this paper will therefore need to acquire observations one to two weeks prior to peak optical light, a non-trivial task.  

This soft X-ray emission, which has not been predicted by analytic models or captured in previous simulations, poses both an opportunity and a challenge for future TDE parameter estimation.  It challenges X-ray continuum fitting models, which estimate SMBH parameters by modeling thermal soft X-rays as having a disk origin \cite{Lin+18, Wen+20, MummeryBalbus20}.  Our results suggest that the earliest X-ray emission, especially pre-peak of the optical/UV light curve, should at least sometimes be excluded from disk modeling.  However, this early-time X-ray emission is surprisingly bright, and if it can be observed, may offer an excellent probe of both the circularization process and of ``nuisance parameters'' such as viewing angle.  We again caution that, as the X-ray emission is produced by hot gas well above $R_{\rm th}$, its quantitative details may be sensitive to our approximate procedures for producing synthetic spectra.  A fuller investigation of the potential and problems related to this early-time X-ray peak will have to be deferred to a full parameter study of circularization simulations.

\section{Comparison to Past Simulations}
\label{sec:comparison}

As was summarized in the main text, past global hydrodynamical simulations have addressed the highly challenging problem of TDE debris evolution via one of three approximations:
\begin{enumerate}
    \item Reducing the dynamic range by using an intermediate-mass black hole rather than a supermassive one\cite{RamirezRuizRosswog09, Guillochon+14, Shiokawa+15}.
    \item Reducing the dynamic range by using an eccentric stellar orbit rather than a parabolic one\cite{Hayasaki+13, Hayasaki+16, Bonnerot+16, Sadowski+16, Liptai+19}.
    \item Reducing the resolution requirements by not tracking stream evolution directly, but instead assuming a prescription for mass injection at the self-intersection point\cite{BonnerotLu20, Bonnerot+21} or elsewhere\cite{Curd21}.
\end{enumerate}
Our results make none of these idealizations or unlikely parameter choices.  Furthermore, the simulations presented in this paper are among the first to evolve TDE debris in radiation hydrodynamics, following a few recent works\cite{Hayasaki+20, Bonnerot+21, Curd21} in accounting for dynamical aspects of radiation fields.  The greater realism of our simulations comes at the cost of runtime, as we are only able to evolve the debris for the first $\approx 65$ days post-disruption.  The only other work we are aware of to avoid approximations (i-iii) above is the recent simulation of Andalman et al.\cite{Andalman+22} in general relativistic hydrodynamics, which likewise is only able to cover the first few days of post-disruption debris evolution.  The major caveat to our work is its lack of accurate treatment of gravity, which only approximates the leading-order effect of general relativity and, more importantly is smoothed over a scale of $60 R_\odot$ for computational tractability.  As a secondary caveat, we do not include magnetohydrodynamics (MHD), although past simulations have found that MHD has little impact on the circularization process\cite{Sadowski+16, Curd21}.  In this section, we compare our results to past numerical hydrodynamics simulations of TDEs.

At early times, our results appear most consistent with approximation (1), which was investigated most thoroughly in the paper of Shiokawa et al.\cite{Shiokawa+15}.  As in that work, we see early-time dissipation that is dominated by the compression shocks caused by the pericentric nozzle.  Dissipation at the self-intersection shock is always sub-dominant, in contrast to a range of other simulations that find minor dissipation at the nozzle and a great deal of early-time dissipation at the stream self-intersection point\cite{Hayasaki+13, Hayasaki+16, Bonnerot+16, Liptai+19, Curd21, Andalman+22}.  Likewise, at early times (before $t\approx 55$ days), our simulation exhibits very inefficient circularization and a low rate of mass loss in outflows (Fig. \ref{fig:circ}).  These overall dynamical features are most consistent with past simulations in approximation (1).

However, as time proceeds, our results appear to more closely resemble those made with artificial mass injection schemes, i.e. approximation (3), as well as the recent work of Andalman et al. -- though with one key difference.  Starting around $t\approx 55$ days, our results show evidence for significant outflows, disagreeing with both IMBH simulations\cite{Shiokawa+15} as well as some low-$e$ TDE simulations\cite{Hayasaki+13, Hayasaki+16, Bonnerot+16}, but in qualitative agreement with many of the most realistic simulations to date\cite{Sadowski+16, BonnerotLu20, Andalman+22}.  Likewise, the rapid growth of circularization efficiency beginning around $t \approx 55~{\rm d}$ differs notably from what is seen in IMBH TDE simulations.  Equivalently, the gas eccentricities seen in our simulation (ED Fig. \ref{fig:ecc_history}) reach lower values at fixed radius than is the case in approximation (1)\cite{Shiokawa+15}.  As detailed earlier, the late-time growth in circularization efficiency is largely driven by interactions between returning streams and an incompletely circularized, non-axisymmetric disk\cite{Andalman+22}, and naturally leads to runaway circularization.  These stream-disk interactions are analogous to but distinct from the ``secondary shocks'' seen in simulations using artificial mass injection schemes\cite{BonnerotLu20, Bonnerot+21}, which are generated by a wide-angle spray of matter leaving the self-intersection point before interacting with the nascent accretion flow (in our case, stream-disk interactions come from dense, cold streams penetrating deeper into the equivalent flow).  

The weaker circularization efficiency in IMBH TDE simulations\cite{Guillochon+14, Shiokawa+15} may be due to differences in the stream-disk interaction geometry.  We speculate that the larger aspect ratio of streams produced by an IMBH disruption leads to stream-disk dissipation at larger dimensionless ($r/R_{\rm g}$) radii, while the smaller aspect ratio of SMBH TDE streams allows them to penetrate farther in to a nascent accretion flow and dissipate energy more efficiently.  The key difference between our results and those of many recent SMBH TDE simulations is the unimportance of the self-intersection shock in our work.  Dissipation at the self-intersection point kick-starts a process of runaway circularization in Andalman et al.\cite{Andalman+22}, and is by construction the initially dominant dissipation mechanism in artificial mass injection simulations\cite{BonnerotLu20, Bonnerot+21}.  However, dissipation due to stream self-intersection is always subdominant in our simulation, as can be seen by eye in Fig. \ref{fig:projection}, or in quantitative detail in \S \ref{app:circ} of this Methods.  Another interesting contrast with TDE simulations using mass injection schemes is that we do not see the formation of a retrograde accretion flow\cite{BonnerotLu20}; the inner accretion flow in our simulation inherits the orientation of its angular momentum from its parent star.

The qualitative agreements and disagreements described here can partially be attributed to the lack of strong relativistic precession in our simulation, which is absent due to our choice of pericenter ($R_{\rm p}/R_{\rm g} \approx 47$ is only a weakly relativistic pericenter, at least in terms of the large self-intersection radius it produces).  Small apsidal precession angles weaken the self-intersection shock and seem to eliminate its ability to generate outflows (see e.g. the lack of strong outflows in the weakly relativistic ``simulation B'' of Bonnerot et al\cite{Bonnerot+21}.  Dissipation at the self-intersection point is further weakened in our work by the mismatch between densities in outgoing and ingoing debris streams, as we showed in \S \ref{app:converge} and ED Fig. \ref{fig:convergence}.  More puzzling, however, is the strength of the shocks produced by vertical compression.  Analytic arguments \cite{Guillochon+14} suggest that these shocks should be strong in IMBH TDEs (as is seen in simulations\cite{Shiokawa+15}) but weak in SMBH TDEs, as we have simulated here.  While we defer a full analysis for future work, it appears that the kinetic energy of vertical stream collapse has been somewhat augmented by our realistic equation of state.  The additional heating injected into the dynamically cold streams by recombination puffs them up, allowing them to accelerate to greater vertical collapse velocities by the time of pericenter return and increasing the kinetic energy budget available for thermalization in shocks (Appendix \ref{app:stream}).

In summary, our simulation of an astrophysically typical TDE, performed without any mass injection onto the grid, has found some entirely novel phenomena in TDE debris evolution, both hydrodynamical (efficient circularization in a TDE with a weakly relativistic pericenter; Fig. \ref{fig:circ}) and directly observable (an early-time soft X-ray transient; Fig. \ref{fig:obs}).  We have also reproduced many individual findings of past approximate simulations: early-time dominance of the compression shock\cite{Shiokawa+15}, general dominance of shock power over accretion power\cite{Shiokawa+15} (though we note again that by the end of the simulation, our upper limit on unsimulated accretion power has become comparable to the simulated shock power), substantial early-time outflows\cite{Sadowski+16, BonnerotLu20, Andalman+22}, and runaway circularization from stream-disk interactions\cite{Andalman+22}.  Of the three existing approximations for studying TDE debris evolution mentioned at the start of this section, all contain important qualitative differences with our results, though also some important qualitative similarities.  At early times, our broad results most closely resemble those of approximation (1), while at late times there is some qualitative similarity to approximation (3).  Approximation (2) appears quite different in the limit of very low stellar eccentricity, although the qualitative resemblance increases in past simulations with $e \gtrsim 0.95$.  

We conclude that truly accurate predictive models must be made with parabolic orbits, realistic MBH masses, and a simulation grid that self-consistently tracks stellar debris at all times.  While this is a formidable challenge, our paper shows that it is achievable at least up to the point of peak light with modern moving-mesh hydrodynamic codes.  Such predictive models are urgently needed at this time of rapid advance in time-domain observational astronomy, and will help near-future samples of hundreds to thousands\cite{BenAmi+22, BricmanGomboc20} of optical/UV TDEs maximize their scientific potential.


\end{methods}


\begin{addendum}
 \item[Data Availability Statement] The simulation results presented in this study are available from the corresponding author upon request. \textsc{RICH} is  publicly available to download from \href{https://gitlab.com/eladtan/RICH}{https://gitlab.com/eladtan/RICH}.  Animated versions of the simulation results are available online, including both \href{https://www.youtube.com/watch?v=O3IWCPO_Thk}{an overall, top-down view of the disruption and formation of an accretion flow}, and also \href{https://www.youtube.com/watch?v=_kpBQjwf3Z4}{a side view that better illustrates the formation of an outflow}.
  \item[Author Information] 
 Correspondence and requests for materials should be addressed to ES~\\(elad.steinberg@mail.huji.ac.il).  \blue{Both authors declare no competing interests (financial or otherwise).}

\end{addendum}

\renewcommand{\figurename}{Extended Data Figure}
\setcounter{figure}{0}

\begin{figure}
\centering
\begin{tabular}{cc}
(a) \includegraphics*[width=8.0cm]{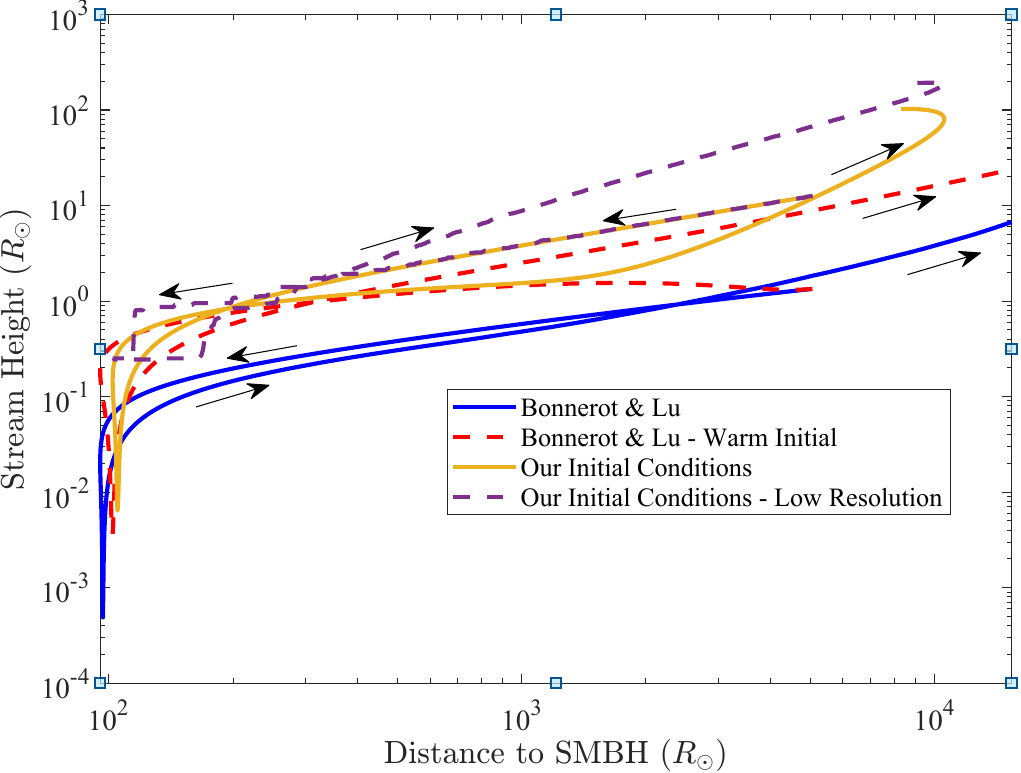} &
(b) \includegraphics*[width=8.0cm]{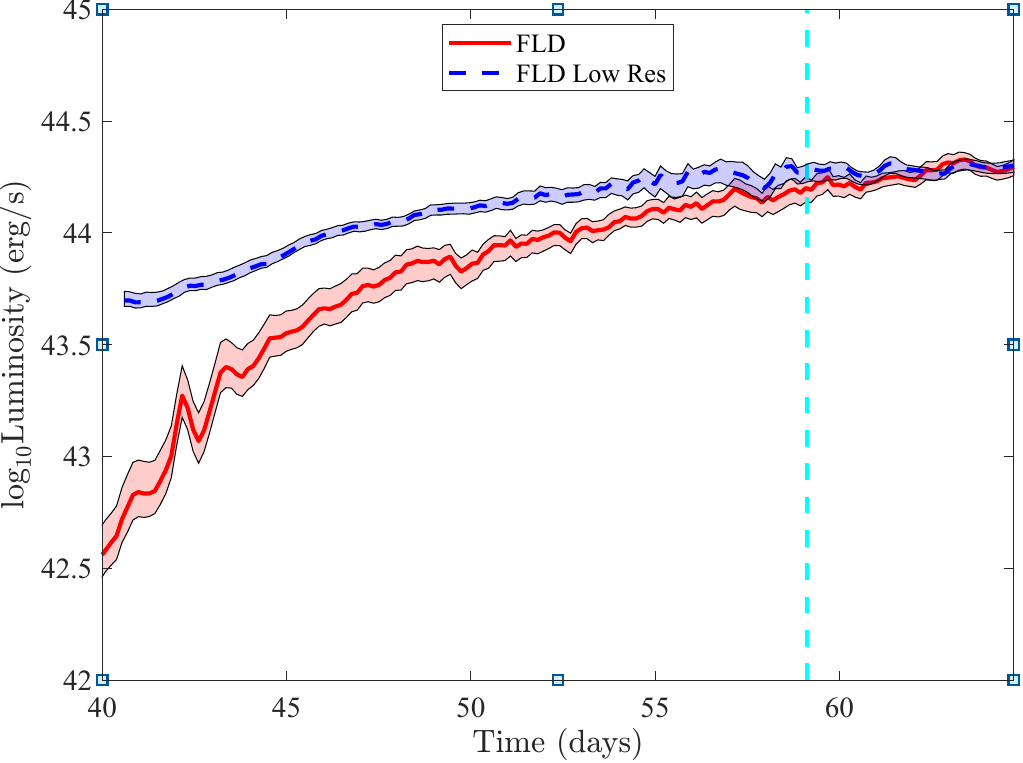} 
\end{tabular}
\caption{{\bf Convergence checks for our simulation.} Panel (a) shows stream scale-heights in 2D simulations designed to study the sensitivity of stream compression to numerical resolution and input physics.  The blue solid line reproduces the study of BL22\cite{BonnerotLu22} from their dynamically cold initial conditions and adiabatic EOS.  The red dashed curve uses a more realistic initial temperature of 4000K, but is otherwise the same. The solid yellow line shows high-resolution 2D results with the initial conditions taken from our 3D simulation (as well as our realistic EOS). The dashed purple line is the same as the yellow line, but with a reduced vertical resolution that corresponds to our 3D resolution. Arrows show the direction of the temporal evolution. In all of the lines, the streams are evolved until they reach the self-intersection radius for the first time.  Panel (b) shows the bolometric light curve from our fiducial 3D run (red curves) and a low-resolution 3D test run with $20\%$ the total number of grid cells.  Excellent convergence is achieved well before the fallback time (vertical dashed line).  Shaded regions correspond to different viewing angles.}
\label{fig:convergence}
\end{figure}


\begin{figure}
\centering
\begin{tabular}{cc}
(a) \includegraphics*[width=8.0cm]{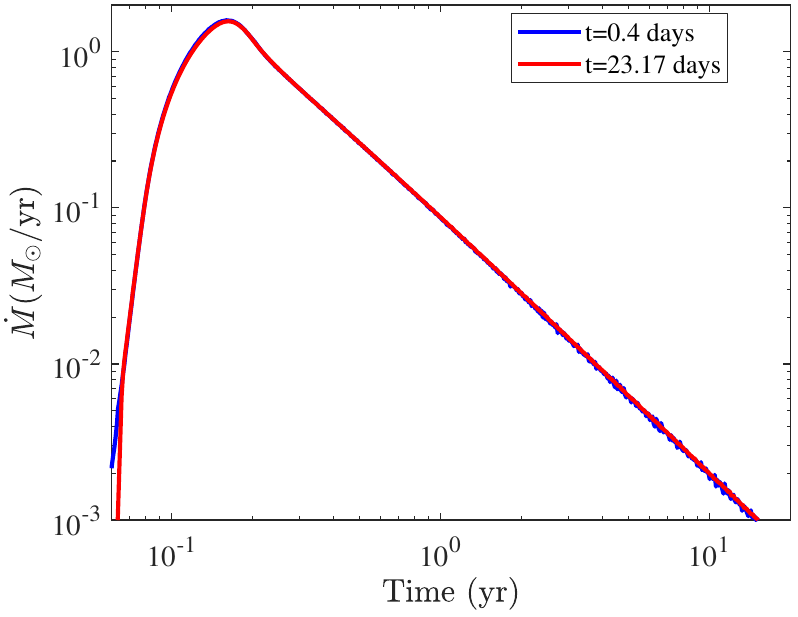} &
(b) \includegraphics*[width=8.0cm]{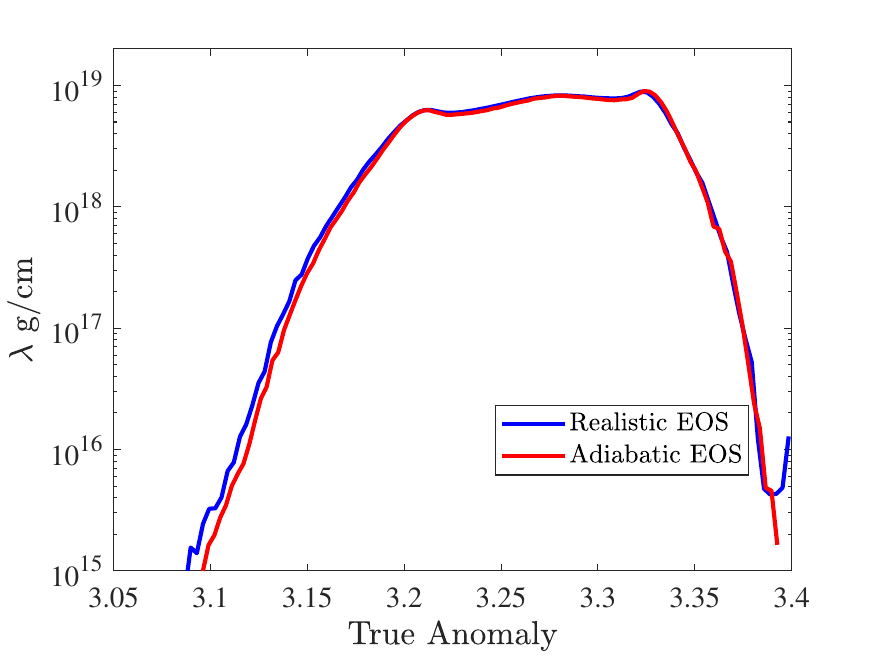} \\
(c) \includegraphics*[width=8.0cm]{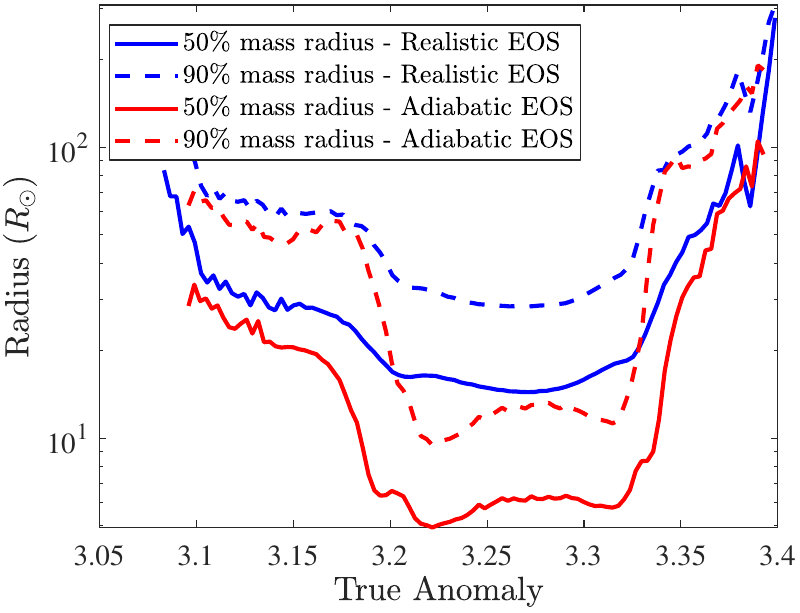} 
\end{tabular}
\caption{{\bf Hydrodynamic and geometric features of the bound debris streams.}  Panel (a) shows the mass fallback rate $\dot{M}_{\rm fb}(t)$ at two different epochs: one immediately post-disruption (blue) and another after the most tightly bound debris has passed apocenter for the first time (red).  The overlap between the curves illustrates the relatively minor effect of recombination on center-of-mass orbital dynamics, as well as the lack of gravitational fragmentation in the debris streams.  In panel (b), we show the stream linear density $\lambda$ as a function of orbital true anomaly $f$ (in the Newtonian limit, $f=0$ is pericenter, and $f=\pi$ apocenter).  As expected, there is little difference between the adiabatic (red) and realistic (blue) equations of state.  In panel (c), we plot the cylindrical radii in the transverse direction that enclose $50\%$ (solid) and $90\%$ (dashed) of the stream mass (colors are the same as in panel b).  Here we see that heating effects from recombination can puff up bound debris streams at the factor $\approx 3$ level in comparison to an adiabatic stream, leading to more violent recompression shocks at pericenter.}
\label{fig:lambda_radius}
\end{figure}

\begin{figure}
\centering
\includegraphics*[width=8.0cm]{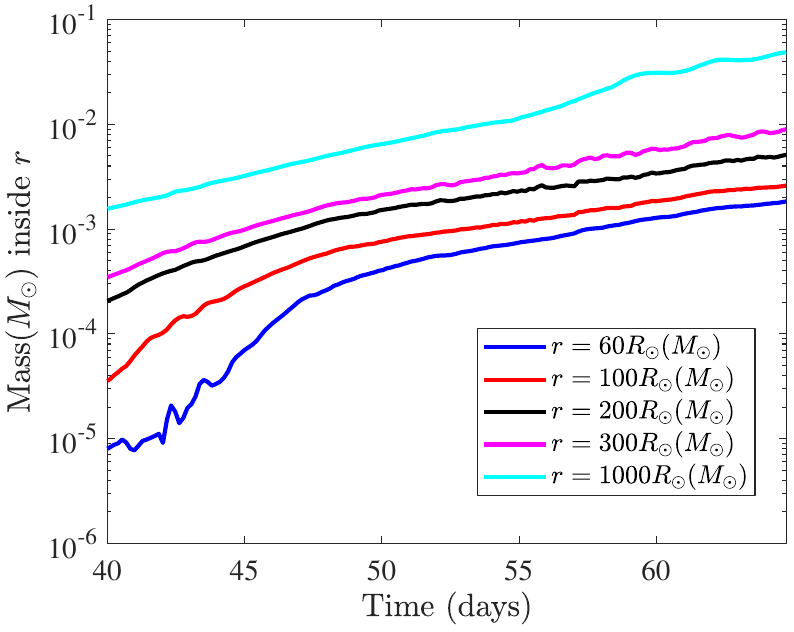}
\caption{{\bf Mass profiles.}  The instantaneously enclosed mass in the inner radii of the simulation, plotted as a function of time since disruption. For the entire duration of the simulation, the fraction of stellar debris inside the gravitational softening length of $h=60 R_\odot$ never exceeds 
$\approx 1.5 \times 10^{-3}$ (blue curve). The mass fractions inside even larger radial cuts are rather small as well, justifying our use of a softened potential on scales $\lesssim h$. }
\label{fig:Mr}
\end{figure}

\begin{figure*}
\begin{tabular}{cc}
\includegraphics*[width=8.2cm]{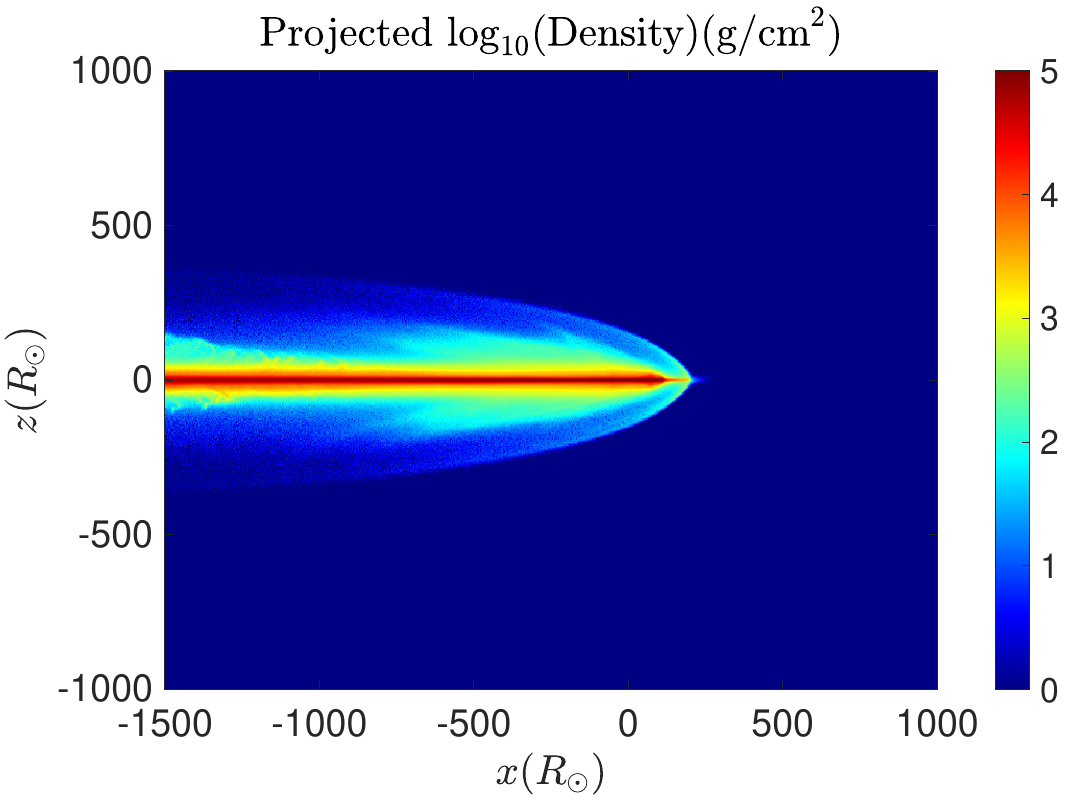} & 
\includegraphics*[width=8.2cm]{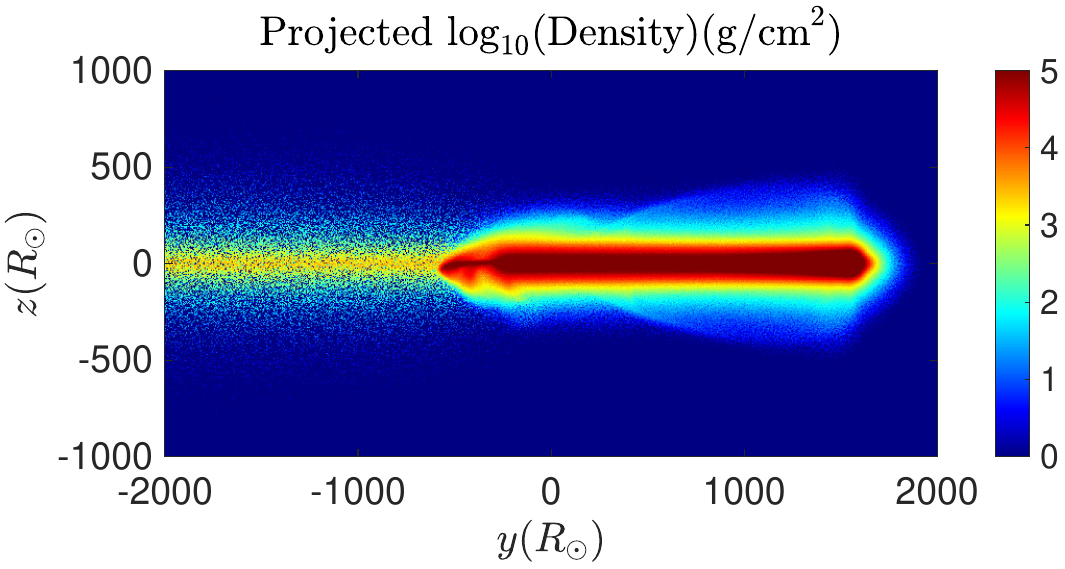} \\
\includegraphics*[width=8.2cm]{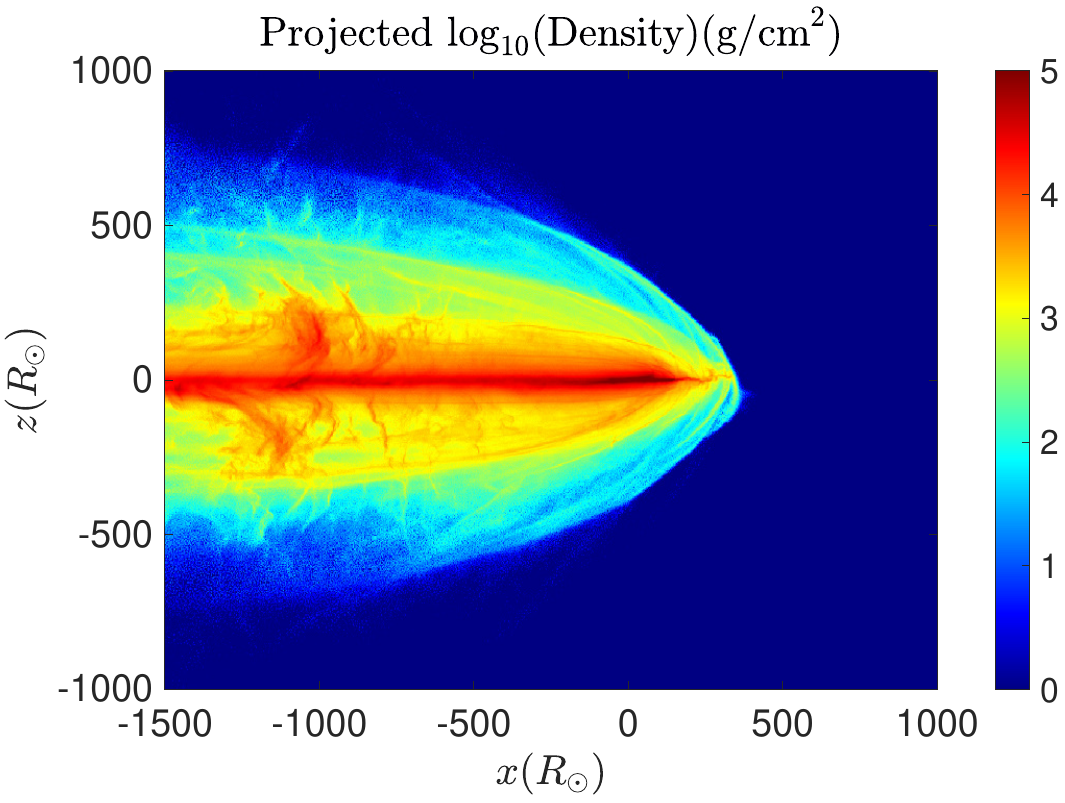} & 
\includegraphics*[width=8.2cm]{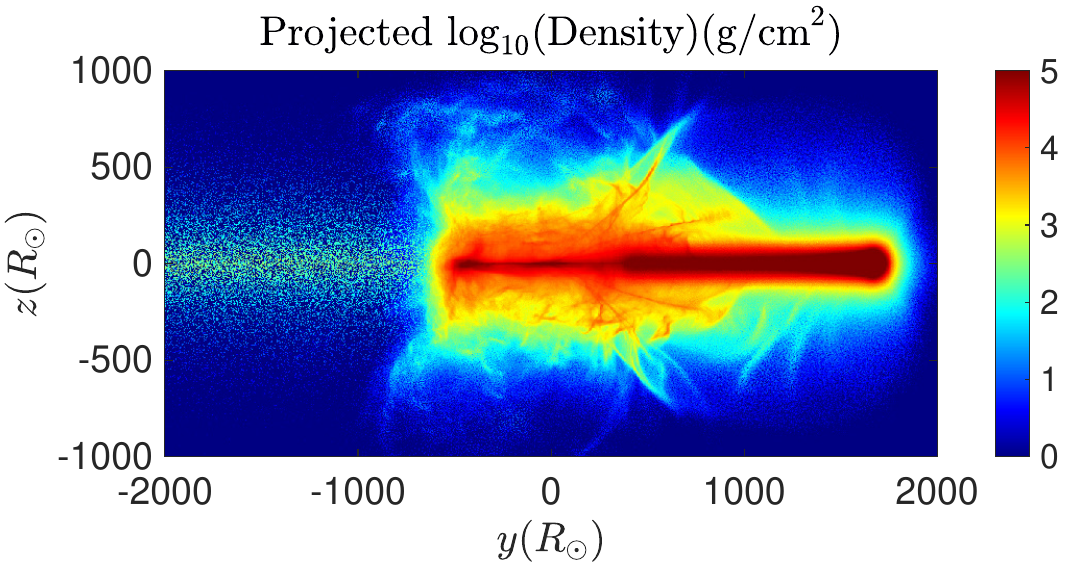} \\
\includegraphics*[width=8.2cm]{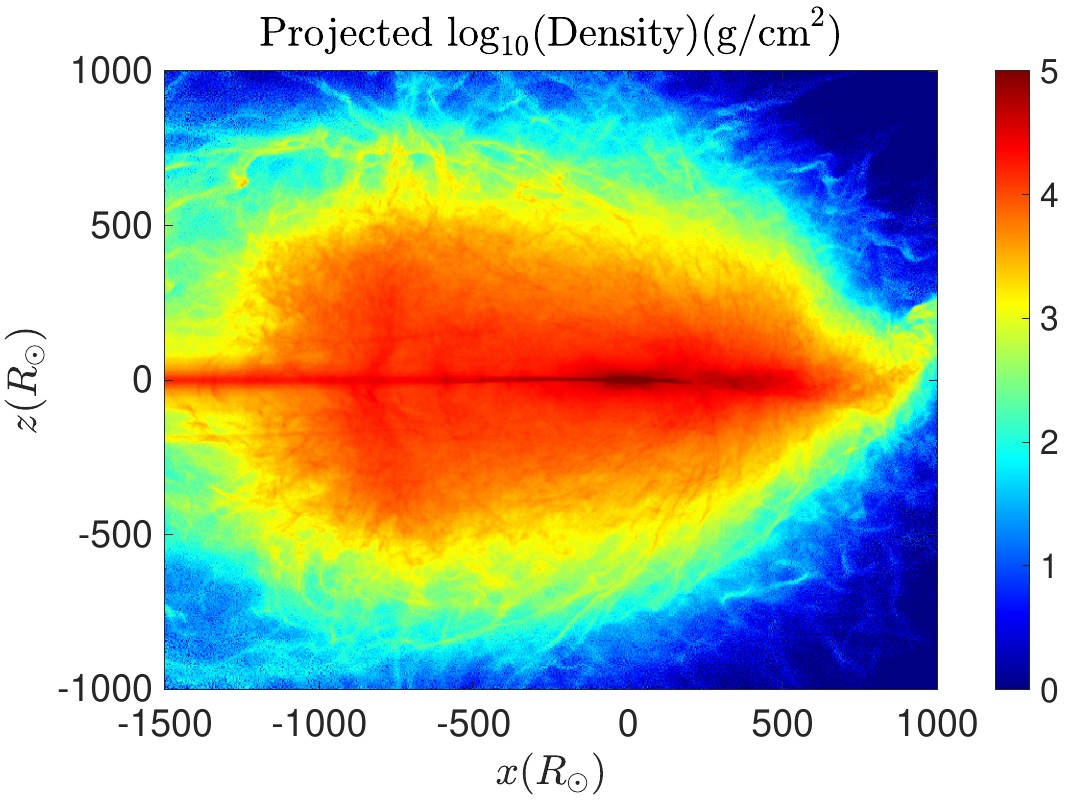} & 
\includegraphics*[width=8.2cm]{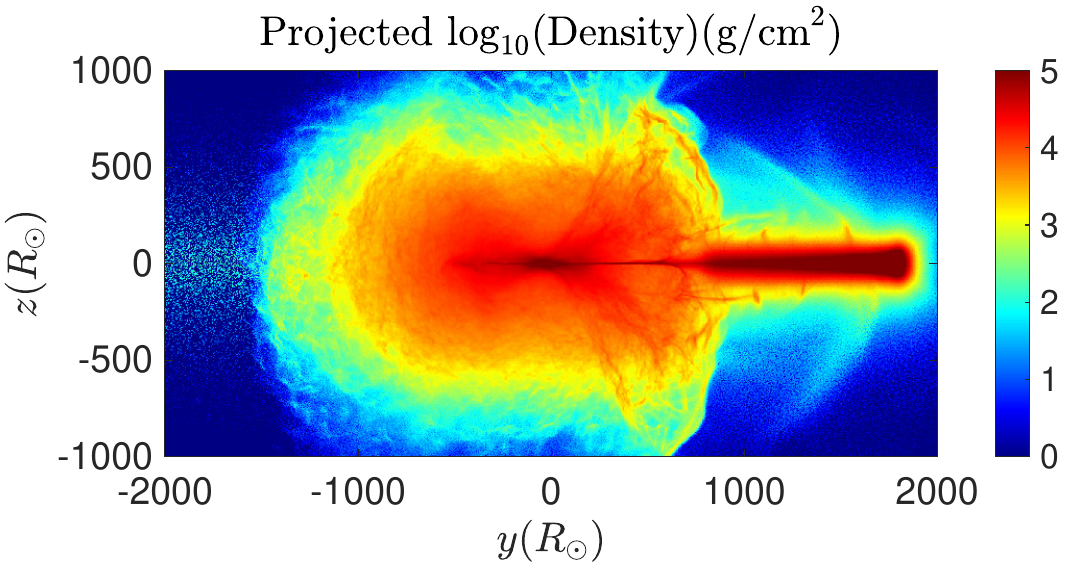} \\
\end{tabular}
\caption{{\bf Density profiles.}  Projections of gas density $\rho$ onto the $xz$ (left) and $yz$ (right) planes, shown at $t=47$ (top), $t=55$ (middle), and $t=62$ (bottom) days.  Logarithmic color schemes are labeled in sidebars.  Early on, most dynamically bound debris has not yet returned to pericenter, and remains in geometrically thin debris streams.  By peak light (e.g. the $t=62$ day panels), an order unity fraction of the bound debris has undergone partial circularization, inflating into a quasi-ellipsoidal accretion flow on scales of $\sim 1000 R_\odot$.}
\label{fig:densityGeometry}
\end{figure*}

\begin{figure*}
\begin{tabular}{cc}
\includegraphics*[width=8.5cm]{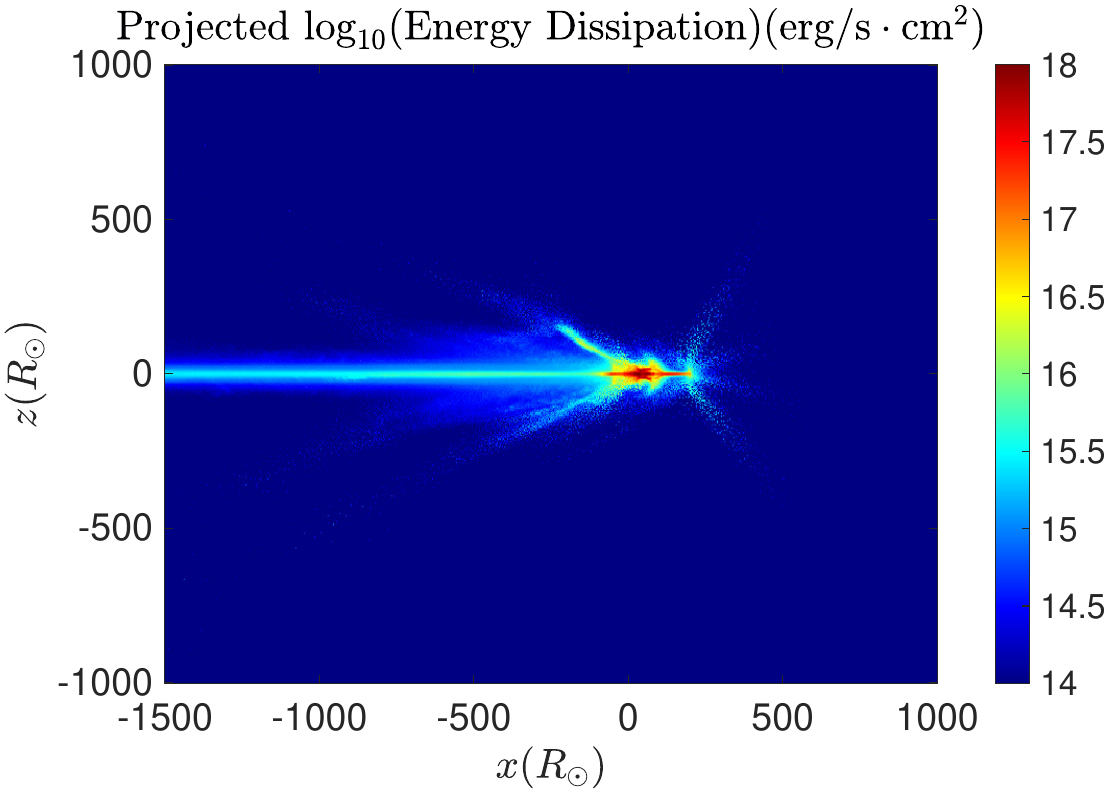} & 
\includegraphics*[width=8.5cm]{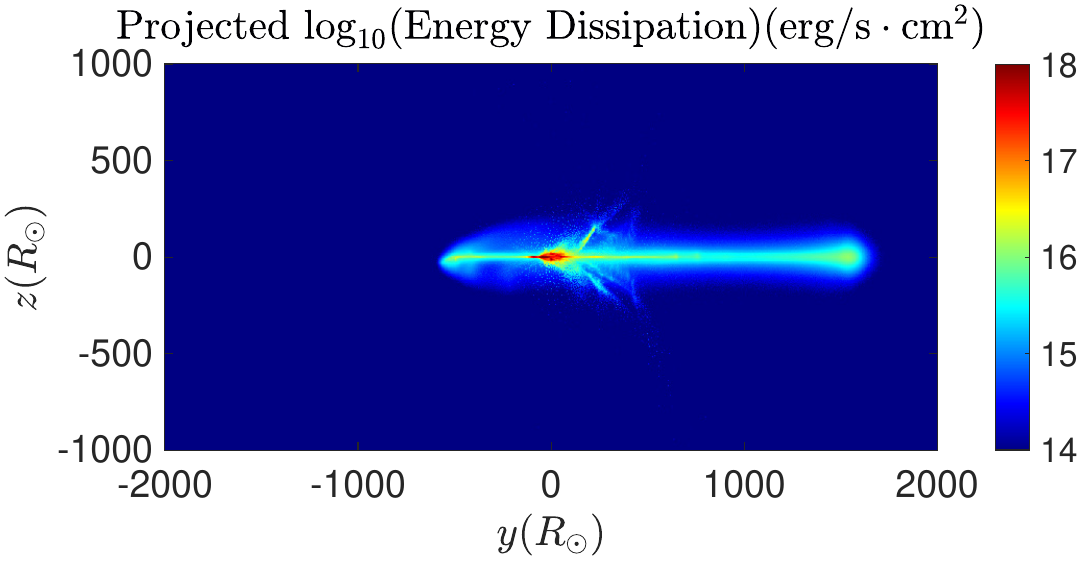} \\
\includegraphics*[width=8.5cm]{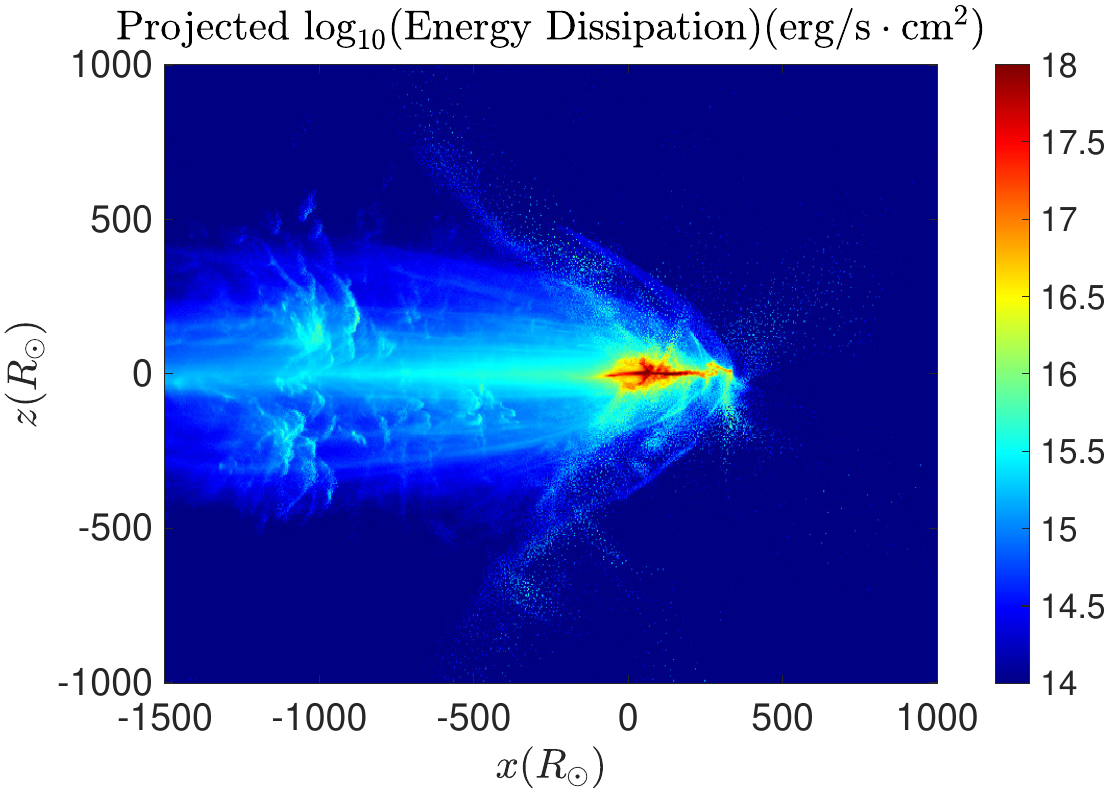} & 
\includegraphics*[width=8.5cm]{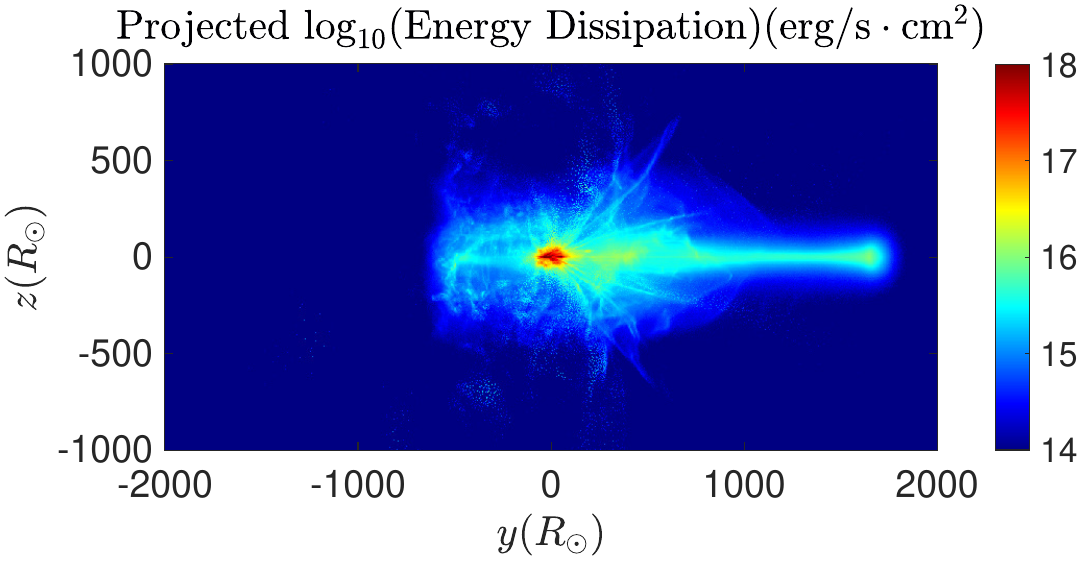} \\
\includegraphics*[width=8.5cm]{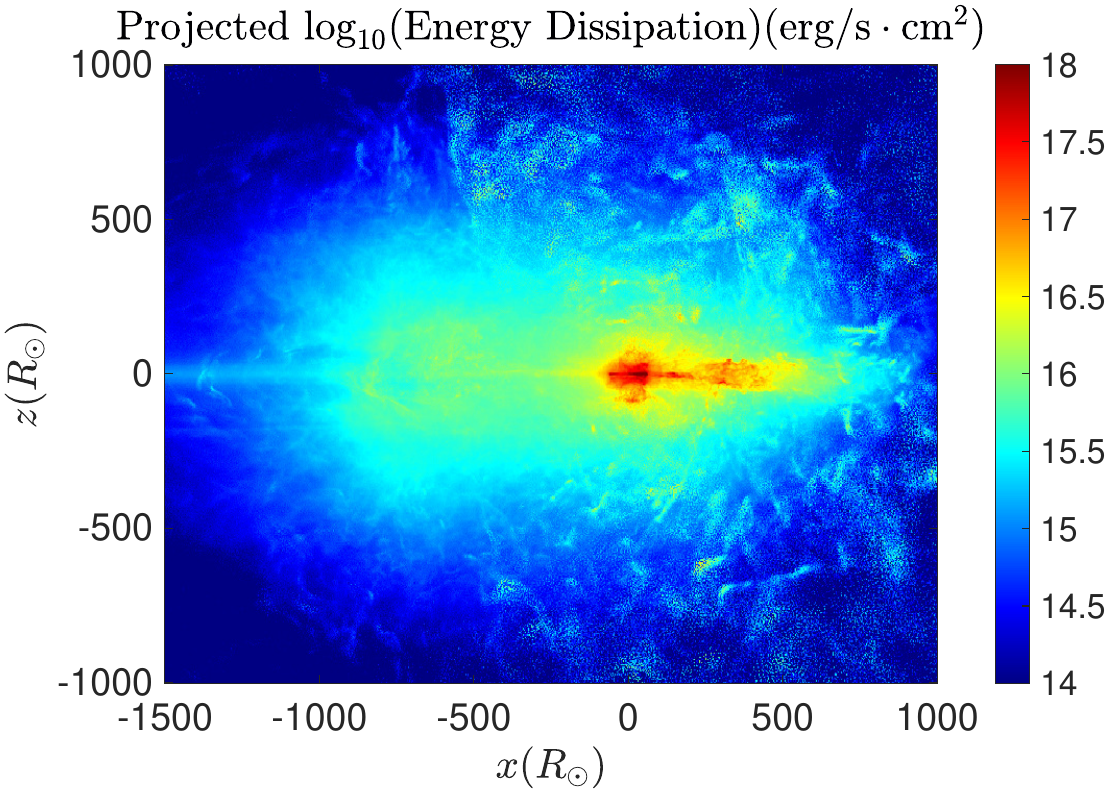} & 
\includegraphics*[width=8.5cm]{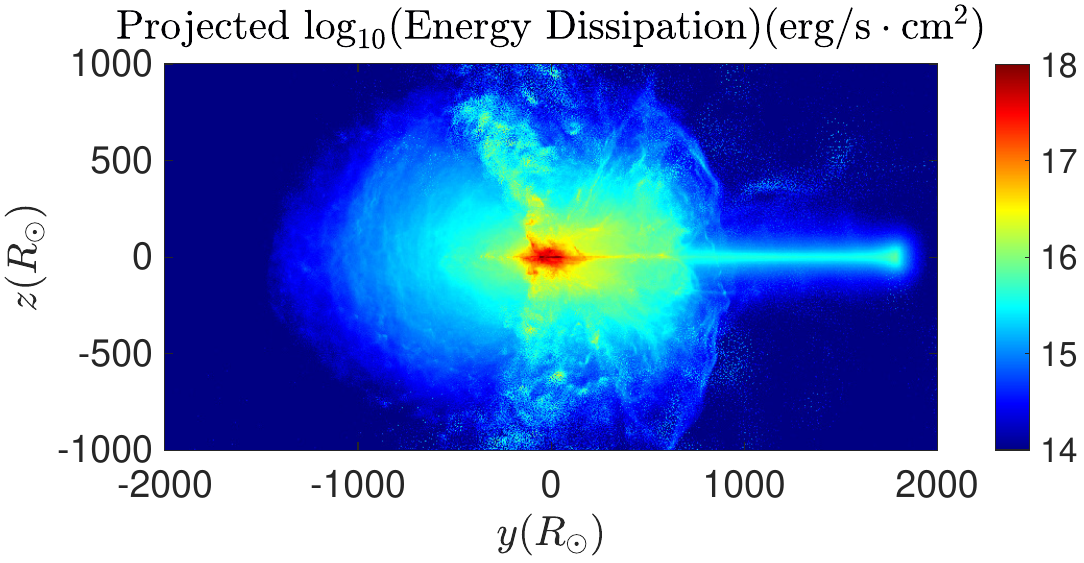} \\
\end{tabular}
\caption{{\bf Dissipation rates.}  Projections of energy dissipation rates $\dot{u}$ onto the $xz$ (left) and $yz$ (right) planes, shown at $t=47$ (top), $t=55$ (middle), and $t=62$ (bottom) days. Logarithmic color schemes are labeled in sidebars.  At early times, dissipation is localized near pericenter in the nozzle shock. Later, as a quasi-ellipsoidal accretion disk forms, the locus of dissipation extends to larger radii at stream-disk shocks.
}
\label{fig:dissipationGeometry}
\end{figure*}

\begin{figure*}
\centering
\begin{tabular}{cc}
\includegraphics*[width=8.0cm]{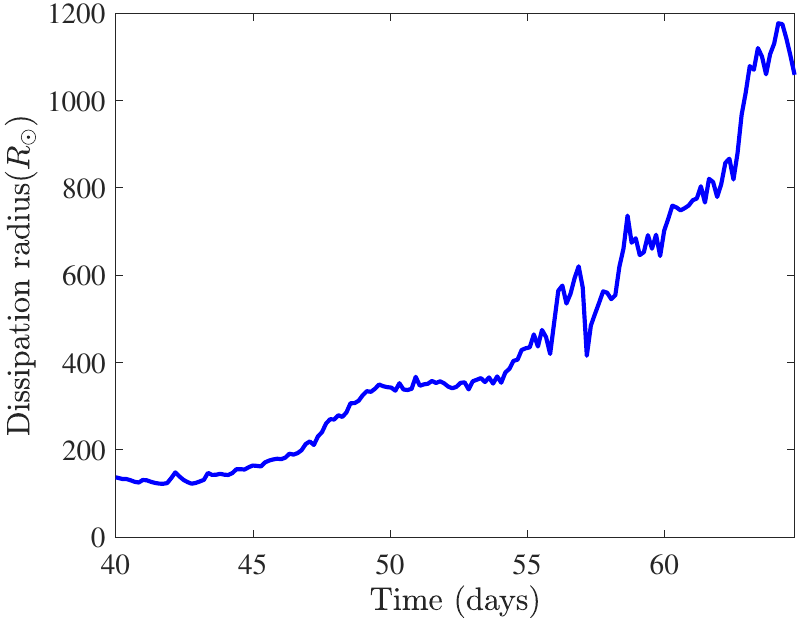} \\
\includegraphics*[width=8.0cm]{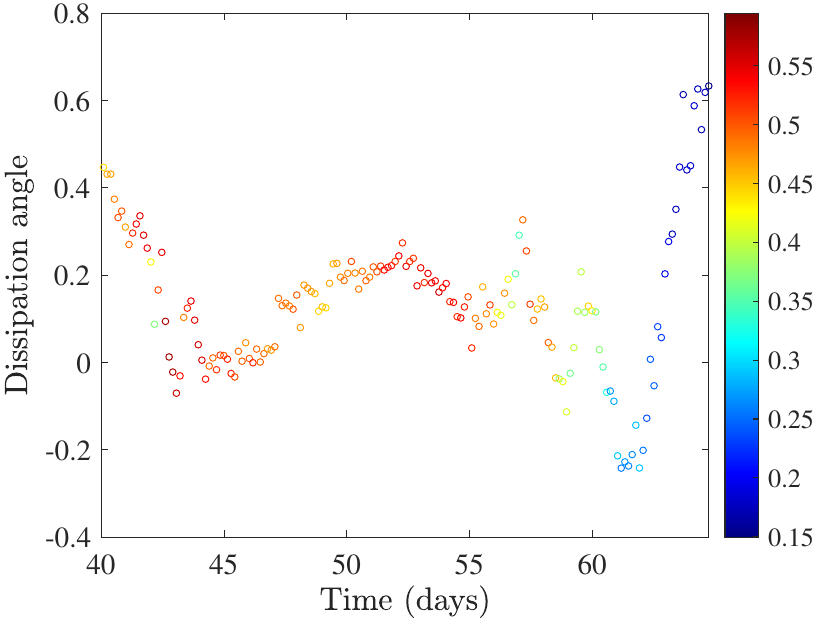}
\end{tabular}
\caption{{\bf Time evolution of the spatial geometry of dissipation.}  Panel (a): the dissipation radius $r_{\rm diss}=\int r \dot{E}dV/\int\dot{E}dV$.  At early times, the primary site of dissipation is at small radii near stream pericenter.  The dissipation radius grows slowly until $\approx 55$ days, when it begins to increase in size at a much faster rate.  Panel (b): time evolution of the average azimuthal dissipation angle $\phi_{\rm diss}$.  The color bar measures the dimensionless angular localization of dissipation, $R_{\rm diss}\le 1$.  Both are calculated via the formulae ${\bf R}_{\rm diss} =\int \hat{\bf r} \dot{E}dV/\int\dot{E}dV$ and  $\cos\phi_{\rm diss} = {\bf R}_{\rm diss} \cdot \hat{\bf x} / |R_{\rm diss}|$.  At early times, the primary site of dissipation is localized at positive angles near and slightly downstream of stream pericenter (i.e. in material that is free-falling into its compression).  The dissipation angle begins to oscillate strongly around and after $t\approx 55$ days, and strongly delocalizes in angle.}
\label{fig:dissipationR}
\end{figure*}

\begin{figure}
\centering
\includegraphics*[width=8.0cm]{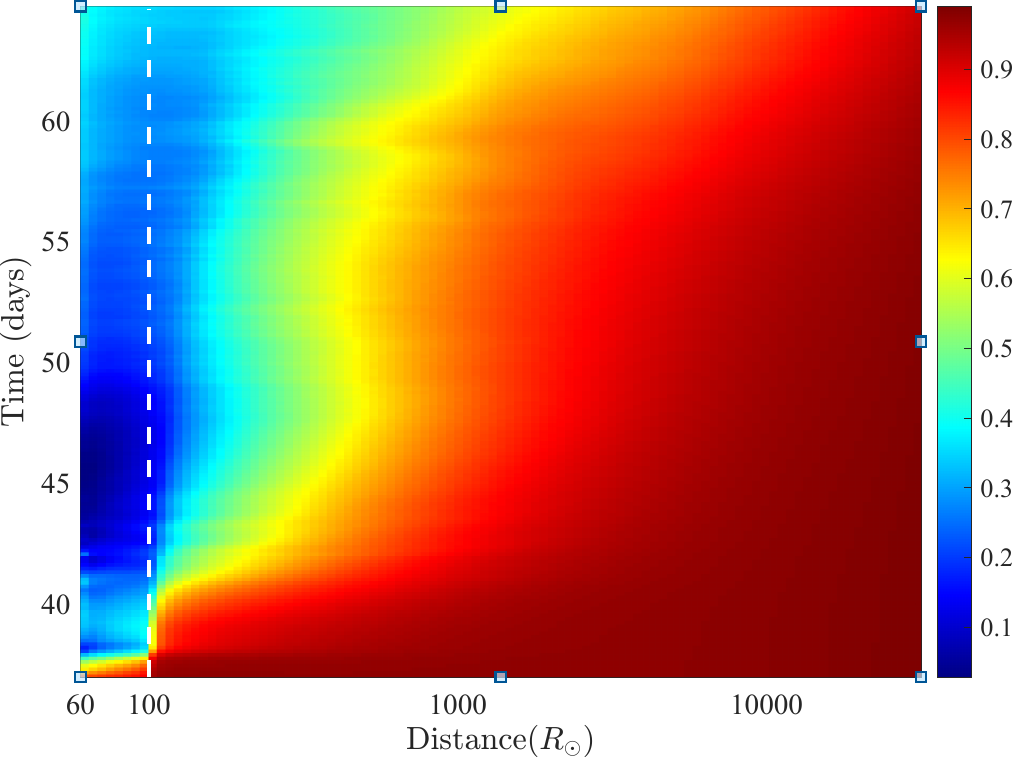}
\caption{{\bf Time evolution of gas eccentricity.}  The mass-weighted eccentricity $e$ (color-coded via the legend on the right) of the bound gas as a function of distance from the SMBH.  The vertical dashed line shows the original tidal radius, and the left side of the plot terminates at the softening radius.  At all radii, $e$ generally decreases with time, reaching low, quasi-circular values $\lesssim 0.3$ inside a radius of $100-200 R_\odot$ by the end of the simulation.}
\label{fig:ecc_history}
\end{figure}

\begin{figure*}
\centering
\begin{tabular}{cc}
\includegraphics*[width=8.0cm]{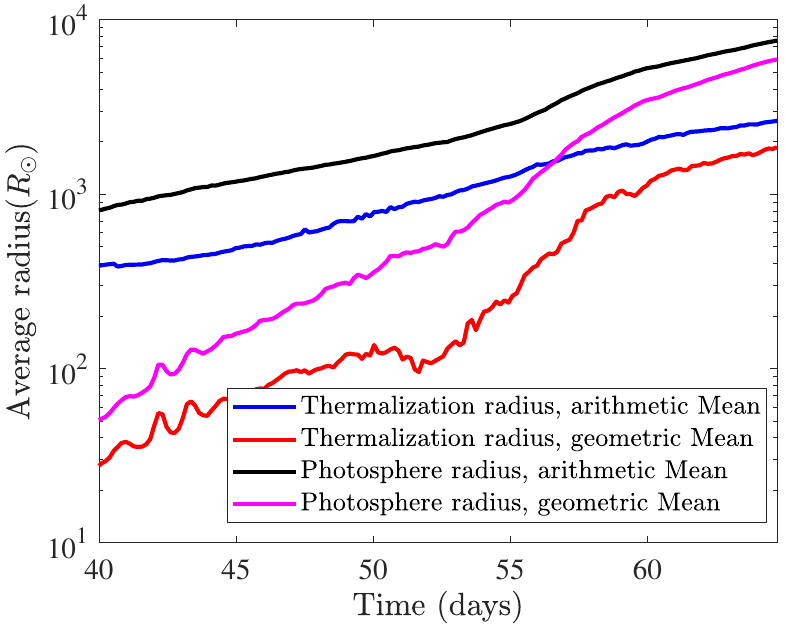} \\
\includegraphics*[width=8.0cm]{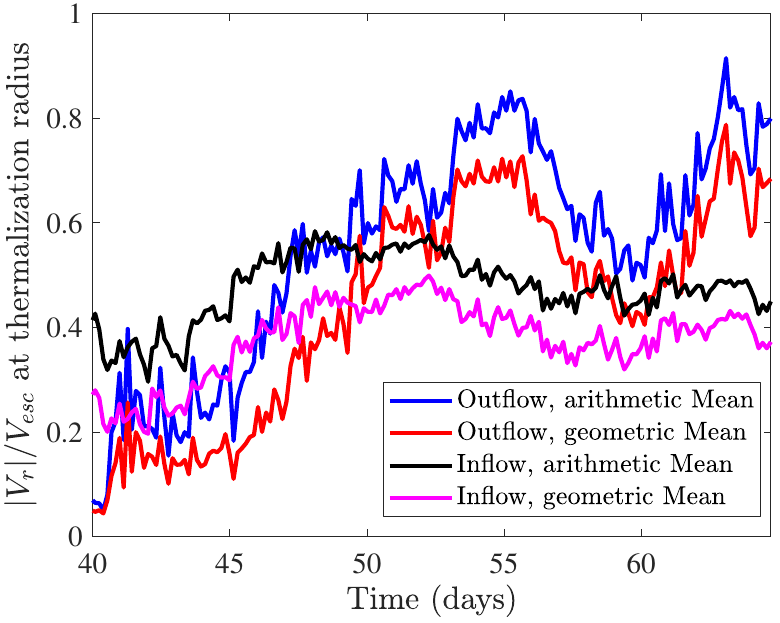}
\end{tabular}
\caption{{\bf Features of the emission surfaces.}  Panel (a): angle-averaged thermalization ($R_{\rm th}$; blue, red curves) and photospheric ($R_{\rm ph}$; black, pink curves) radii plotted as functions of time.  The photosphere is always larger than $R_{\rm th}$, demonstrating the dominance of scattering opacity.  This is shown with an arithmetic mean (blue, black curves) and a geometric mean (red, pink curves).  The stronger time evolution in the geometric means highlights the higher initial asymmetry in the actual thermalization surface/photosphere, but the convergence of the two means shows that this asymmetry declines over time.  Panel (b): the dimensionless gas radial velocity, $|V_{\rm r}|/V_{\rm esc}$, at the thermalization radius, plotted as a function of time.  As in Fig. \ref{fig:rTherm}, we show both arithmetic (blue, black) and geometric (red, pink) angle averages.  We break up the thermalization surface into inflowing (black, pink) and outflowing (blue, red) components.  The ratio $V_{\rm r}/V_{\rm esc} < 1$ always, implying that the gas on the thermalization surface is, on average, bound to the SMBH.}
\label{fig:rTherm}
\end{figure*}


\begin{figure}
\centering
\begin{tabular}{cc}
(a)\includegraphics*[width=8.0cm]{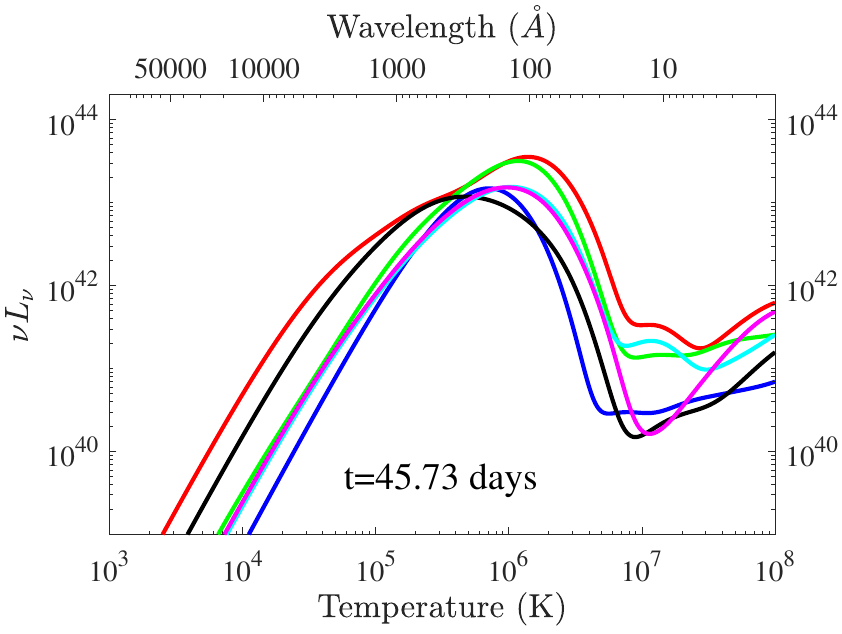} \\
(b)\includegraphics*[width=8.0cm]{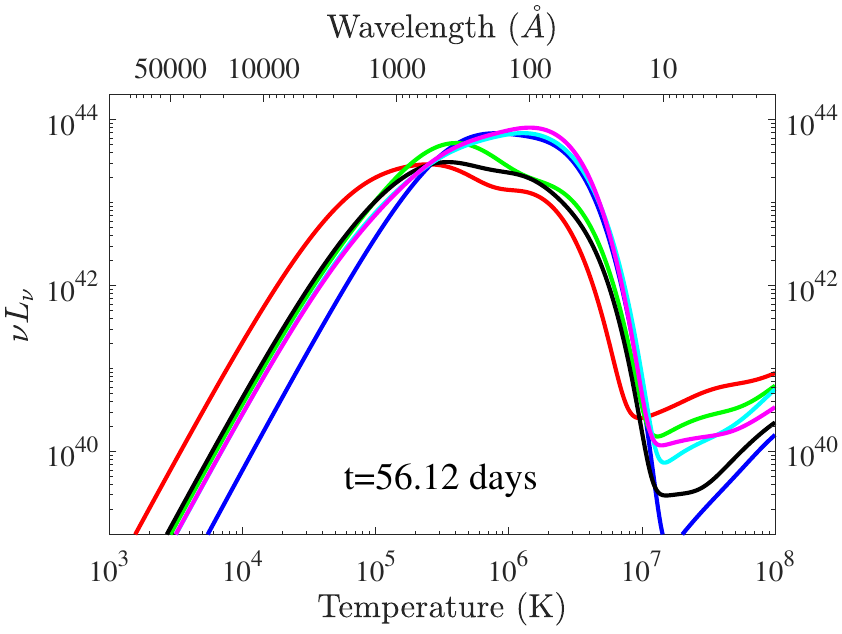} \\
(c)\includegraphics*[width=8.0cm]{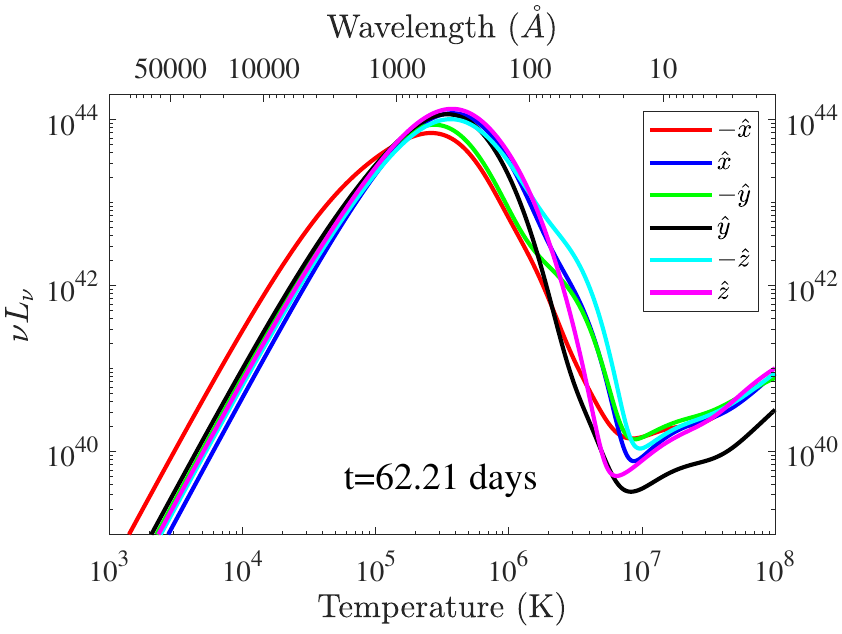} \\
\end{tabular}
\caption{{\bf Synthetic spectra.}  The emitted spectrum $\nu L_\nu$ plotted against wavelength $\lambda$ (and equivalently effective temperature $T$) at 3 different times: $t=45.73$ days ({\it panel a}), $t=56.12$ days ({\it panel b}), and $t=62.21$ days ({\it panel c}).  In each of these three panels, the six different curves correspond to six different viewing angles on the simulation's principal axes.  The stellar debris orbits in the $xy$-plane with a pericenter on the positive $x$-axis. As time goes on, the emitted spectra become cooler.}
\label{fig:nuLnu}
\end{figure}

\begin{figure}
\centering
(a)\includegraphics[width=8.0cm]{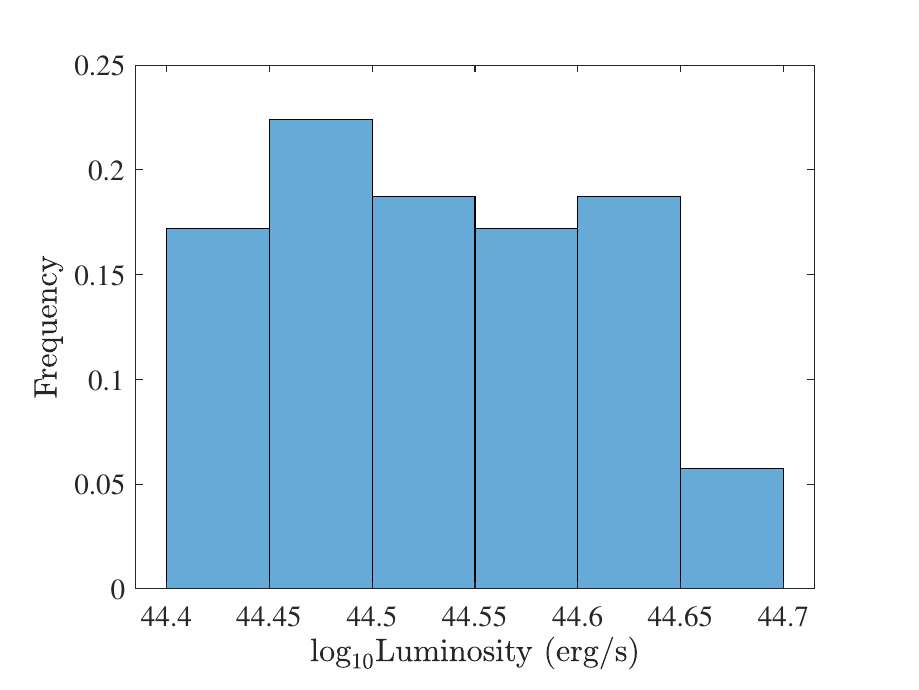}
(b)\includegraphics[width=8.0cm]{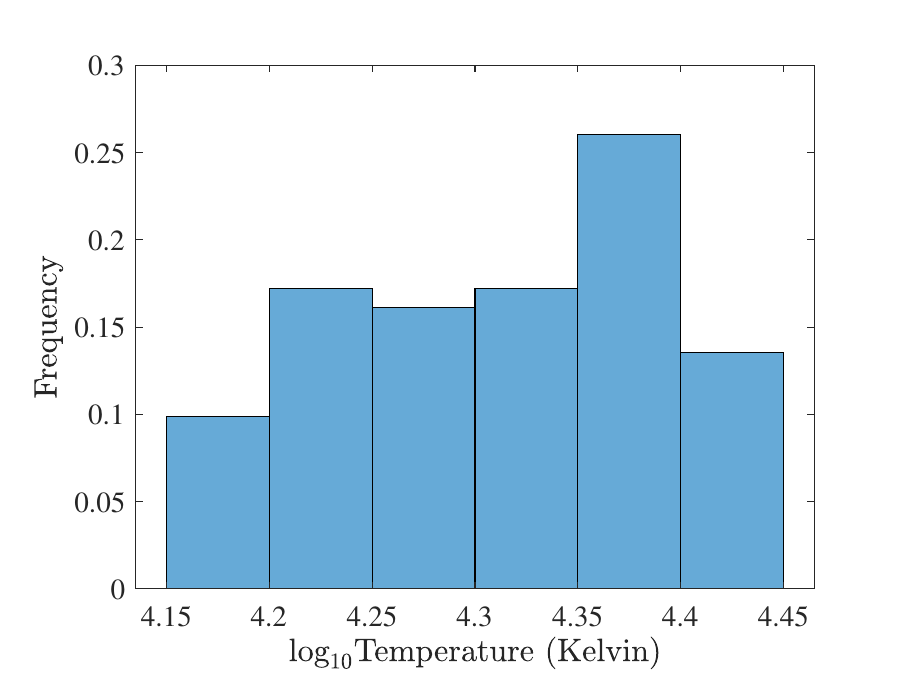}
(c)\includegraphics[width=8.0cm]{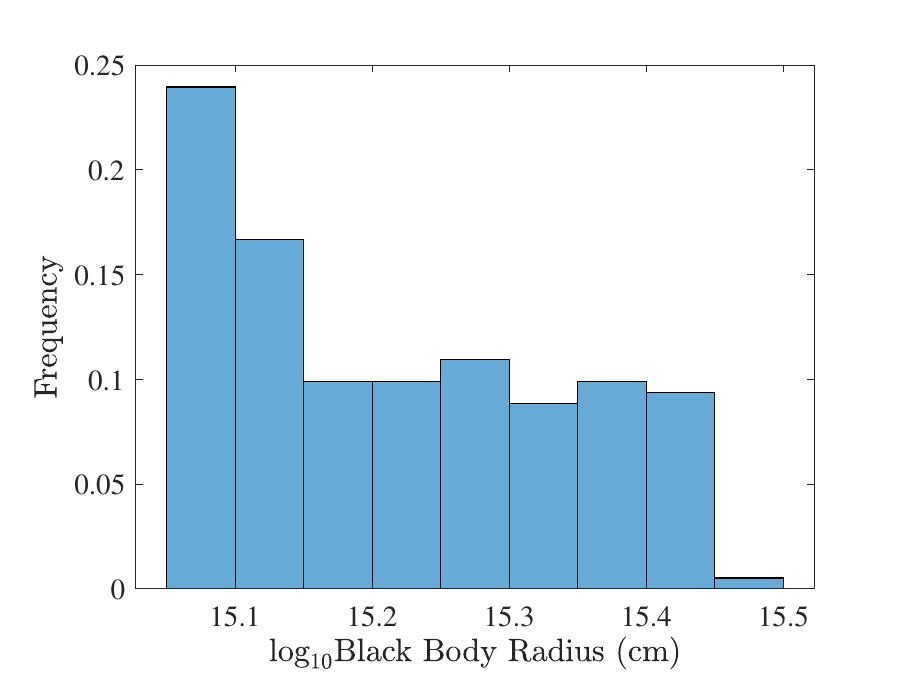}
\caption{{\bf Properties of optical/UV black body fits.}  The occurrence rates for black body properties of our simulation at $t=65$ days, binned over 192 different observing angles.  The top, middle, and bottom panels show occurrence rates for fitted luminosity, blackbody temperature, and blackbody radius, respectively.  There is a limited spread (roughly half an order of magnitude) in each of these quantities, depending on viewing angle.}
\label{fig:T_hist}
\end{figure}

\begin{figure}
\centering
(a)\includegraphics[width=8.0cm]{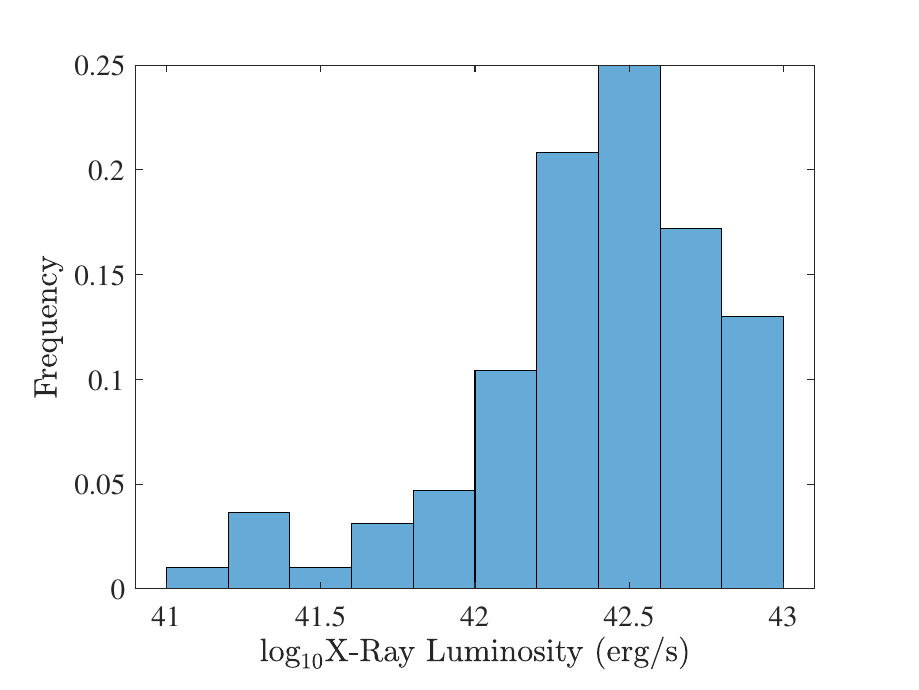}
(b)\includegraphics[width=8.0cm]{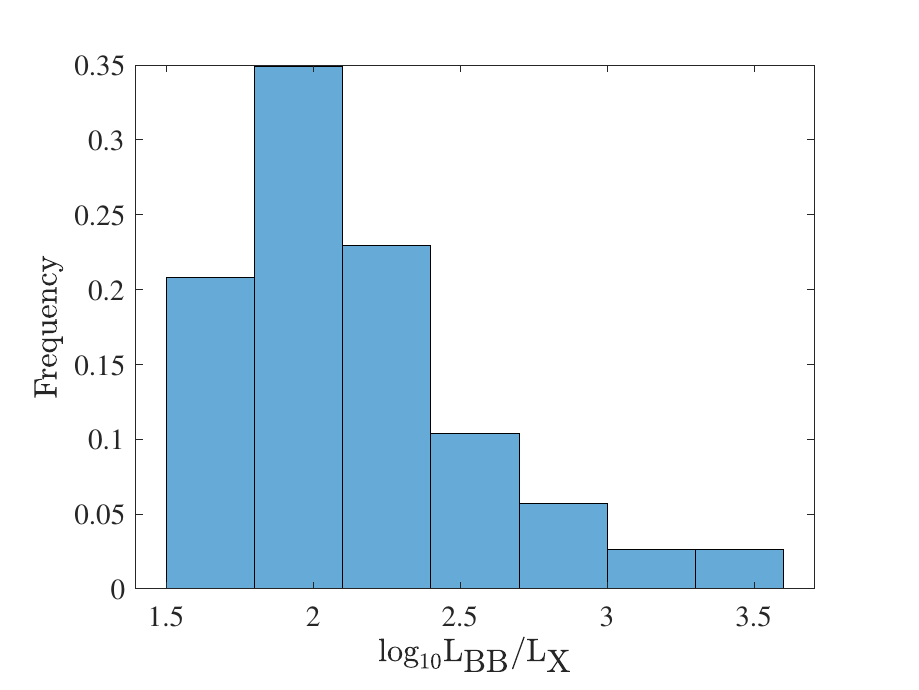}
\caption{{\bf Thermal X-ray properties.}  The occurrence rates for X-ray luminosities in our simulation at $t=65$ days, binned over 192 different observing angles.  The top and bottom panels show occurrence rates for {\it Swift} XRT-band X-ray luminosity and the optical-to-X-ray ratio, respectively.  Typical X-ray luminosities vary by an order of magnitude depending on viewing angle, from $10^{42-43} ~{\rm erg~s}^{-1}$, with some outlying viewing angles as dim as $10^{41}~{\rm erg~s}^{-1}$.  By the time of peak optical light shown here, the optical to X-ray ratio is $\sim 100$, with more than an order of magnitude of scatter.}
\label{fig:X_hist}
\end{figure}

\end{document}